\documentclass[12pt]{article}
\usepackage{amsmath,amssymb}
\usepackage{bm}
\usepackage{color}
\usepackage{cite}
\usepackage{mathtools}

\usepackage{here}

\def\slash#1{\not\!\!#1}

\begin{document}

\title{
\begin{flushright}
\ \\*[-80pt]
\begin{minipage}{0.2\linewidth}
\normalsize
EPHOU-20-013\\*[50pt]
\end{minipage}
\end{flushright}
{\Large \bf
Classification of three generation models by orbifolding magnetized $T^2 \times T^2$
\\*[20pt]}}

\author{
Kouki Hoshiya,
~Shota Kikuchi,
~Tatsuo Kobayashi, \\
 Yuya Ogawa, and
~Hikaru Uchida
\\*[20pt]
\centerline{
\begin{minipage}{\linewidth}
\begin{center}
{\it \normalsize
Department of Physics, Hokkaido University, Sapporo 060-0810, Japan} \\*[5pt]
\end{center}
\end{minipage}}
\\*[50pt]}

\date{
\centerline{\small \bf Abstract}
\begin{minipage}{0.9\linewidth}
\medskip
\medskip
\small
We study orbifolding by the $\mathbb{Z}_2^{\rm (per)}$ permutaion of  $T^2_1 \times T^2_2$ with magnetic fluxes and its twisted orbifolds.
We classify the possible three generation models which lead to non-vanishing Yukawa couplings on the magnetized $T^2_1 \times T^2_2$ and orbifolds including the $\mathbb{Z}_2^{\rm (per)}$ permutation and $\mathbb{Z}_2^{\rm (t)}$ twist.
We also study the modular symmetry on such orbifold models.
As an illustrating model, we examine the realization of quark masses and mixing angles.
\end{minipage}
}

\begin{titlepage}
\maketitle
\thispagestyle{empty}
\end{titlepage}

\newpage


\section{Introduction}
\label{Intro}

The Standard Model (SM) was established by the discovery of the Higgs particle.
It turns out to be consistent with almost all the observations on experiments so far.
However, in particle physics there still remain several mysteries, e.g. the origin of the gauge group configuration $SU(3)_C \times SU(2)_L \times U(1)_Y$, the origin of the flavor structure such as the mass hierarchy and the flavor mixing and so on.
Many approaches have been ever proposed to understand these issues.
Among them, extra dimensional theories such as superstring theory are attractive, 
because Yukawa couplings are written by the wavefunction overlap integral among quarks, leptons and the Higgs field 
in the extra dimensional space and they can be hierarchically suppressed depending their wavefunction behaviors. 
In particular, superstring theory is a promising candidate for the unified theory and it predicts 
the ten-dimensional (10D) space-time, i.e. the extra six-dimensional compact space in addition to our four-dimensional (4D) space-time.
Such a compact space may be an origin of the flavor structure, i.e. the number of generations, the mass hierarchy and the 
flavor mixing.

When we start with extra dimensional theories such as superstring theory, 
it is an important keypoint how to derive 4D chiral theory, because 
the SM is a chiral theory.
On the other hand, the torus compactification is one of simplest compactifications, but 
it leads to 4D non-chiral theory.

The torus compfactification with magnetic fluxes leads to 4D chiral theories 
\cite{Bachas:1995ik,Blumenhagen:2000wh,Angelantonj:2000hi,Blumenhagen:2000ea}.
Furthermore, the number of zero-modes, i.e. the generation number,  is determined by the magnitude of fluxes.
In addition, zero-mode profiles are quasi-localized, and they can lead to hierarchical Yukawa couplings \cite{Cremades:2004wa}.
Thus, the torus compactification with magnetic fluxes are quite interesting.
Moreover, in Refs.~\cite{Kobayashi:2018rad,Ohki:2020bpo,Kikuchi:2020frp,Kikuchi:2020nxn}  
it was shown that their flavor structures are controlled by the modular symmetry \footnote{See also Ref.~\cite{Kobayashi:2018bff}.}.
Recently, the modular flavor symmetries were studied extensively from the phenomenological 
viewpoint of 4D model building \cite{Feruglio:2017spp}
\footnote{See for recent works in heterotic orbifold models Refs.~\cite{Baur:2019kwi,Nilles:2020tdp}.}.

The toroidal orbifold is another way to realize a 4D chiral theory \cite{Dixon:1985jw}.
The twisted orbifold models with magnetic fluxes were studied \cite{Abe:2008fi,Abe:2013bca,Abe:2014noa}.
The adjoint scalar fields, i.e. the open string moduli, are projected out in orbifold models with magnetic fluxes.
In addition, the number of zero-modes is different from one in the torus compactification with magnetic fluxes 
\cite{Abe:2008fi,Abe:2013bca,Abe:2014noa,Kobayashi:2017dyu,Sakamoto:2020pev}.
Then, various models can be constructed compared with the simple torus compactification.
Indeed, the three generation models have been classified \cite{Abe:2008sx,Abe:2015yva}.
Realistic quark and letpon masses as well as their mixing angles and CP phase were studied 
\cite{Abe:2012fj,Abe:2014vza,Fujimoto:2016zjs,Abe:2016eyh,Kobayashi:2016qag}.
In addition, shifted orbifolding was also studied \cite{Fujimoto:2013xha}.

For factorizable higher dimensional tori, e.g. $T^4=T^2_1\times T^2_2$ and 
$T^6=T^2_1\times T^2_2 \times T^2_3$ and their orbifolds, the flavor structure is a tensor product of 
the flavor structures of $T^2_i$. 
The three generations can be realized in the models, where three zero-modes appear on only one of $T^2_i$ and 
only one zero-mode appears on the other tori 
\footnote{See for non-factorizable  fluxes Refs.~\cite{Antoniadis:2009bg,Abe:2013bba}.}.
If three zero-modes for both left-handed and right-handed quarks appear on the same $T^2_i$, e.g. $T^2_1$, 
the flavor structure such as the mass ratios, mixing angles and CP phase is determined by 
wavefunctions on $T^2_1$.
Wavefunctions on the other $T^2_i$ contribute to only the overall factor of Yukawa matrices.
If three zero-modes for left-handed quarks appear on the torus, e.g. $T^2_1$ and three zero-modes 
for right-handed quarks appear on a different torus, e.g. $T^2_2$, 
their Yukawa matrices can be written by $Y_{ij}=a_ib_j$.
Their ranks are equal to one, and two generations of quarks are massless.
That is not realistic.

Recently the orbifolding by the $\mathbb{Z}_2^{\rm (per)}$ permutation between $T^2_1$ and $T^2_2$ 
was studied \cite{Kikuchi:2020nxn}.
Either symmetric modes or anti-symmetric modes under $\mathbb{Z}_2^{\rm (per)}$ are projected out.
The flavor structure in these models is not a simple tensor product, but 
leads to a new and rich flavor structure.
Our purpose in this paper is to classify three generation models by 
such permutation orbifolding of $T^2\times T^2$ and its twsited orbifolds.

In this paper, we study the orbifolding by the $\mathbb{Z}_2^{\rm (per)}$ permutation between $T^2_1$ and $T^2_2$ as well as their twisted orbifolds.
We classify the possible three generation models through permutation orbifolding of  the magnetized $T^2 \times T^2$, which lead to non-vanishing Yukawa coupling.
In addition, we classify the three generation models by permutation of the twisted orbifolds.
We also study the modular symmetry of the Yukawa couplings.
We examine an illustrating example towards realization of quark masses and mixing angles.

This paper is organized as follows.
In section \ref{U(N)}, we give a brief review on the torus compactification with magnetic fluxes.
In section \ref{ModonT2xT2}, we review the modular symmetry of the zero-modes on the torus.
In section \ref{Magnetized orbifolds}, we study the orbifolding by permutation of $T^2_1$ and  $T^2_2$.
In section \ref{ThreeonT2xT2}, we classify the possible three generation models through the $\mathbb{Z}_2^{\rm (per)}$ permutation 
orbifolding.
In section \ref{Yukawa}, we study Yukawa couplings and their behavior as the modular forms and give an example of numerical studies on our models.
In section \ref{conclusion}, we conclude this study.
In Appendix, we study the shiftted orbifold models.


\section{$U(N)$ gauge theory on $(T^2)^3$}
\label{U(N)}
First of all, we start with a review of the 10D $U(N)$ super Yang-Mills theory on the background $M^4 \times T^2_1 \times T^2_2 \times T^2_3$ with magnetic fluxes, which is low-energy effective field theory of superstring theory.
(See for details Ref.~ \cite{Cremades:2004wa}.)
Our setup is also applicable as low-energy effective field theory of D7-brane models on $M^4 \times T^2_1 \times T^2_2$. 
The Lagrangian is given as
\begin{align}
  S = \int_{M^4}d^4x \prod_{i = 1,2,3} \int_{T^2_i}dy_{2+2i}dy_{3+2i} \left[-\frac{1}{4g^2}{\rm Tr}(F^{MN}F_{MN})+\frac{i}{2g}{\rm Tr}(\bar{\lambda}\Gamma^MD_M\lambda)\right],
\end{align}
where $M,N=0,...,9$.
Here, $\lambda$ denotes gaugino fields and $F_{MN}$ is the field strength of the vector field $A_M$ written by
\begin{align}
  F_{MN} = \partial_MA_N - \partial_NA_M - i[A_M,A_N].
\end{align}
The covariant derivative $D_M\lambda$ is given by
\begin{align}
  D_M\lambda = \partial_M\lambda-i[A_M,\lambda].
\end{align}
These fields are decomposed into the non-compact space $M^4$ and the compact space $T^2_1 \times T^2_2 \times T^2_3$ as
\begin{align}
  A_M(x,y) &= \sum_{n_1,n_2,n_3} \phi_{M,n_1n_2n_3}(x) \otimes \prod_{i=1,2,3} \phi_{M,n_i}(y_{2+2i},y_{3+2i}) \\
  \lambda(x,y) &= \sum_{n_1,n_2,n_3} \psi_{n_1n_2n_3}(x) \otimes \prod_{i=1,2,3} \psi_{n_i}(y_{2+2i},y_{3+2i}),
\end{align}
where $n_1,n_2,n_3$ are Kaluza-Klein (KK) indices.
We assume the background magnetic fields such as
\begin{align}
  F_{2+2i\ 3+2i} = 2\pi
  \begin{pmatrix}
    M_a^{(i)} \mathbf{1}_{N_a \times N_a} & & \\
    & M_b^{(i)} \mathbf{1}_{N_b \times N_b} & \\
    & & M_c^{(i)} \mathbf{1}_{N_c \times N_c} 
  \end{pmatrix},
  \quad i = 1,2,3,
\end{align}
where $N = N_a + N_b + N_c$ and $M_{\alpha}^{(i)} \in \mathbb{Z}$ ($\alpha = a,b,c)$ to fulfill Dirac's quantization condition on the compact space $T^2$.
Then, the gauge symmetry $U(N)$ is broken to $U(N_a) \times U(N_b) \times U(N_c)$.
For instance, assuming $N_a = 4$, $N_b = 2$ and $N_c = 2$, the Pati-Salam gauge group $SU(4) \times SU(2)_L \times SU(2)_R$ can be realized up to $U(1)$ factors.
These fluxes allow us to choose the vector potentials $A = A_{2+2i}dy_{2+2i}+A_{3+2i}dy_{3+2i}$ like
\begin{align}
	A_{2+2i} &= A_{2+2i}^{\alpha} T_{\alpha} = 0, \\
	A_{3+2i} &= A_{3+2i}^{\alpha} T_{\alpha} \notag \\
	&= 2\pi
	\begin{pmatrix}
		M_a^{(i)}(y_{2+2i}+\zeta^a_{2+2i})\mathbf{1}_{N_a\times N_a} & & \\
		& M_b^{(i)}(y_{2+2i}+\zeta^b_{2+2i})\mathbf{1}_{N_b\times N_b} & \\
		& & M_c^{(i)}(y_{2+2i}+\zeta^c_{2+2i})\mathbf{1}_{N_c\times N_c} \\
	\end{pmatrix},
\end{align}
where $\zeta$ stands for Abelian Wilson line phase on $T^2$, and 
$T_{\alpha}$ are the genarators of remaining gauge symmetry $U(N_a) \times U(N_b) \times U(N_c)$.

For simplicity, we focus on $U(N_a) \times U(N_b)$ part.
Furthermore, we consider vanishing Wilson lines.
The gauge components of the gaugino $\lambda$ are written as
\begin{align}
  \lambda(x,y) = 
  \begin{pmatrix}
    \lambda^{aa}(x,y) & \lambda^{ab}(x,y) \\
    \lambda^{ba}(x,y) & \lambda^{bb}(x,y)
  \end{pmatrix}.
\end{align}
Here, $\lambda^{aa}$ and $\lambda^{bb}$ correspond to the gaugino fields under the unbroken gauge group $U(N_a) \times U(N_b)$, while $\lambda^{ab}$ and $\lambda^{ba}$ are bi-fundamental matter fields, $(N_a,\bar{N}_b)$ and $(\bar{N}_a,N_b)$, under the unbroken gauge group $U(N_a) \times U(N_b)$.
Their 2D spinors on the $i$-th torus
\begin{align}
  \psi_{n_i}(y) = 
  \begin{pmatrix}
    \psi_{n_i+} \\
    \psi_{n_i-}
  \end{pmatrix}, \quad
  \psi_{n_i\pm}(y) = 
  \begin{pmatrix}
    \psi^{aa}_{n_i\pm}(y) & \psi^{ab}_{n_i\pm}(y) \\
    \psi^{ba}_{n_i\pm}(y) & \psi^{bb}_{n_i\pm}(y)
  \end{pmatrix},
\end{align}
are studied in what follows.
Hereafter,
we concentrate on the zero-modes on the torus
and they denote $\psi_{(i)}(y)$.

The torus can be regarded as the complex plane $\mathbb{C}$ divided by a two-dimensional lattice $\Lambda$ spanned by two basis $e_{1(i)}=2\pi R_{i}$ and $e_{2(i)} = 2\pi R_{i} \tau_{i}$, $\tau_{i} \in \mathbb{C}$.
Introducing the complex coordinate on the torus as $z_i = y_{2+2i} + \tau_i y_{3+2i}$, the metric on the torus is given by
\begin{align}
  ds^2 = 2h_{\mu\nu}^{(i)}dz_i^{\mu}d\bar{z}_i^{\nu}, \quad 
  h^{(i)} = |e_{1(i)}|^2
  \begin{pmatrix}
    0 & \frac{1}{2} \\
    \frac{1}{2} & 0
  \end{pmatrix}.
\end{align}
Note that there are identifications $z_i \sim z_i + 1 \sim z_i + \tau_i$ on the torus.
Then, the gamma matrices on the torus are given by
\begin{align}
  \Gamma^{z_i} = \frac{1}{e_{1(i)}}
  \begin{pmatrix}
    0 & 2 \\
    0 & 0
  \end{pmatrix}, \quad
  \Gamma^{\bar{z}_i} = \frac{1}{\bar{e}_{1(i)}}
  \begin{pmatrix}
    0 & 0 \\
    2 & 0
  \end{pmatrix}.
\end{align}
Now, we can write down the Dirac equation for zero-modes on the torus:
\begin{align}
  i\slash{D}\psi_{(i)} = (i\Gamma^{z_i}D_{z_i}+i\Gamma^{\bar{z}_i}D_{\bar{z}_i})\psi_{(i)} = i
  \begin{pmatrix}
    0 & -D^{\dagger} \\
    D & 0
  \end{pmatrix}
  \begin{pmatrix}
    \psi_{(i)+} \\
    \psi_{(i)-}
  \end{pmatrix}
  = 0.
\end{align}
Especially, this equation for $\psi_{(i)+}$ is written as
\begin{align}
  D \psi_{(i)+} &= (\pi R_i)^{-1}(\partial_{\bar{z}_i}\psi_{(i)+}+[A_{\bar{z}_i},\psi_{(i)+}]) \notag \\
  &= (\pi R_i)^{-1}
  \begin{pmatrix}
    \partial_{\bar{z}_i}\psi^{aa}_{(i)+} & (\partial_{\bar{z}_i}+\frac{\pi I_{ab}^{(i)}}{2{\rm Im}\tau_{i}} z_i)\psi^{ab}_{(i)+} \\
    (\partial_{\bar{z}_i}-\frac{\pi I_{ab}^{(i)}}{2{\rm Im}\tau_{i}} z_i)\psi^{ba}_{(i)+} & \partial_{\bar{z}_i}\psi^{bb}_{(i)+}
  \end{pmatrix}
  = 0,
\end{align}
where $A_{\bar{z}_i}$ is defined by $A = A_{z_i}dz_i + A_{\bar{z_i}}d\bar{z}_i$ and $I_{ab}^{(i)}$ denotes $M_a^{(i)}-M_b^{(i)} \neq 0$ satisfying $I_{ab}^{(i)}+I_{ca}^{(i)}=I_{cb}^{(i)}$.
It must be solved under the following boundary conditions on the torus:
\begin{align}
  \begin{array}{c}
    \psi^{aa}_{(i)}(z_i+1) = \psi^{aa}_{(i)}(z_i+\tau_{i}) = \psi^{aa}_{(i)}(z_i), \\
    \psi^{ab}_{(i)}(z_i+1) = e^{i\chi^{ab}_{1(i)}(z_i)} \psi^{ab}_{(i)}(z_i), \quad \psi^{ab}_{(i)}(z_i+\tau_{i}) = e^{i\chi^{ab}_{2(i)}(z_i)} \psi^{ab}_{(i)}(z_i), \\
    \psi^{ba}_{(i)}(z_i+1) = e^{-i\chi^{ab}_{1(i)}(z_i)} \psi^{ba}_{(i)}(z_i), \quad \psi^{ba}_{(i)}(z_i+\tau_{i}) = e^{-i\chi^{ab}_{2(i)}(z_i)} \psi^{ba}_{(i)}(z_i), \\
    \psi^{bb}_{(i)}(z_i+1) = \psi^{bb}_{(i)}(z_i+\tau_{i}) = \psi^{bb}_{(i)}(z_i),
  \end{array} \label{BConTorus}
\end{align}
with
\begin{align}
  \chi^{ab}_{1(i)}(z_i) = \frac{\pi I_{ab}^{(i)}}{{\rm Im}\tau_{i}}{\rm Im}z_i, \quad \chi^{ab}_{2(i)}(z_i) = \frac{\pi I_{ab}^{(i)}}{{\rm Im}\tau_{i}}{\rm Im}(\bar{\tau}_{i}z_i).
\end{align}
As the results, we can find the wavefunctions on the torus such that for $I_{ab}^{(i)}>0$ only $\psi_{(i)+}^{ab}$ and $\psi_{(i)-}^{ba}$ have zero-mode solutions with the degeneracy number $|I_{ab}^{(i)}|$, while for $I_{ab}^{(i)}<0$ only $\psi_{(i)+}^{ba}$ and $\psi_{(i)-}^{ab}$ have such solutions.
Thus,
we can obtain $|I_{ab}^{(i)}|$ generations of  bi-fundamental chiral zero-modes on the magnetized $i$-th torus.
Note that we have one generation and one anti-generation of zero-modes for $I_{ab}^{(i)} = 0$, that is, non-chiral.
The explicit form of the $j^{(i)}$-th zero-mode of $\psi_{(i)\pm}$ is given by
\begin{align}
\psi_{(i)+}^{j^{(i)},|I_{ab}^{(i)}|}(z_i,\tau_{i})
&= \left(\frac{|I_{ab}^{(i)}|}{{\cal A}_{(i)}^2}\right)^{1/4} e^{i\pi |I_{ab}^{(i)}|z_i \frac{{\rm Im}z_i}{{\rm Im}\tau_{i}}} \sum_{l \in \mathbf{Z}} e^{i\pi |I_{ab}^{(i)}|\tau_{i} \left( \frac{j^{(i)}}{|I_{ab}^{(i)}|}+l \right)^2} e^{2\pi i|I_{ab}^{(i)}|z_i \left( \frac{j^{(i)}}{|I_{ab}^{(i)}|}+l \right)} \\
&= \left(\frac{|I_{ab}^{(i)}|}{{\cal A}_{(i)}^2}\right)^{1/4} e^{i\pi |I_{ab}^{(i)}|z_i \frac{{\rm Im}z_i}{{\rm Im}\tau_{i}}}
\vartheta
\begin{bmatrix}
\frac{j^{(i)}}{|I_{ab}^{(i)}|}\\
0
\end{bmatrix}
(|I_{ab}^{(i)}|z_i, |I_{ab}^{(i)}|\tau_{i}), \label{psizero} \\
\psi_{(i)-}^{j^{(i)},|I_{ab}^{(i)}|}(\bar{z}_i,\bar{\tau}_{i})
&= \left(\psi_{(i)+}^{-j^{(i)},|I_{ab}^{(i)}|}(z_i,\tau_{i})\right)^*
\end{align}
where $j^{(i)} = 0,1,...,|I_{ab}^{(i)}|-1$ and ${\cal A}_{(i)} = |e_{1(i)}|^2 {\rm Im}\tau_{i}$ is the area of torus.
The $\vartheta$ function is the Jacobi theta function defined by
\begin{align}
\vartheta
\begin{bmatrix}
a\\
b
\end{bmatrix}
(\nu, \tau)
=
\sum_{l\in \mathbb{Z}}
e^{\pi i (a+l)^2\tau}
e^{2\pi i (a+l)(\nu+b)}.
\end{align}
This function has  the property
\begin{align}
  \vartheta \begin{bmatrix} \frac{r}{N_1} \\ 0 \end{bmatrix} (\nu_1,N_1\tau) \times \vartheta \begin{bmatrix} \frac{s}{N_2} \\ 0 \end{bmatrix} (\nu_2,N_2\tau) 
  &= \sum_{m\in\mathbb{Z}_{N_1+N_2}} \vartheta \begin{bmatrix} \frac{r+s+N_1m}{N_1+N_2} \\ 0 \end{bmatrix} (\nu_1+\nu_2,(N_1+N_2)\tau) \notag \\
  \times \vartheta &\begin{bmatrix} \frac{N_2r-N_1s+N_1N_2m}{N_1N_2(N_1+N_2)} \\ 0 \end{bmatrix} (\nu_1N_2-\nu_2N_1,N_1N_2(N_1+N_2)\tau).
\end{align}
Then, we can find the normalization and product expansions of
the zero-modes \cite{Cremades:2004wa}:
\begin{align}
  &\int_{T^2_i} dz_id\bar{z}_i \left(\psi^{j^{(i)}, |I_{ab}^{(i)}|}_{(i)\pm}(z_i)\right)^* \psi^{k^{(i)},|I_{ab}^{(i)}|}_{(i)\pm}(z_i) 
  = (2{\rm Im}\tau_i)^{-1/2} \delta_{j,k}, \label{Normalization} \\
  &\psi^{j^{(i)},|I_{ab}^{(i)}|}_{(i)\pm} (z_i,\tau_i) \cdot \psi^{k^{(i)},|I_{ca}^{(i)}|}_{(i)\pm} (z_i,\tau_i) = \sum_{m\in\mathbb{Z}_{|I_{cb}^{(i)}|}} Y^{|I_{ab}^{(i)}|,|I_{ca}^{(i)}|,|I_{cb}^{(i)}|}_{j^{(i)},k^{(i)},m}(\tau_i) \psi^{j^{(i)}+k^{(i)}+ m|I_{ca}^{(i)}|,|I_{cb}^{(i)}|}_{(i)\pm} (z_i,\tau_i), \label{ProductExpansion} 
\end{align}
where
\begin{align}
  Y^{| I_{ab}^{(i)}|,|I_{ca}^{(i)}|,|I_{cb}^{(i)}|}_{j^{(i)},k^{(i)},m}(\tau_i) \equiv {\cal A}_{(i)}^{-1/2}\left|\frac{I_{ab}^{(i)}I_{ca}^{(i)}}{I_{cb}^{(i)}}\right|^{1/4} \vartheta
  \begin{bmatrix} \frac{|I_{ca}^{(i)}|j^{(i)}-|I_{ab}^{(i)}|k^{(i)}+|I_{ab}^{(i)}I_{ca}^{(i)}|m}{|I_{ab}^{(i)}I_{ca}^{(i)}I_{cb}^{(i)}|} \\ 0 \end{bmatrix} (0,|I_{ab}^{(i)}I_{ca}^{(i)}I_{cb}^{(i)}| \tau_i) \label{productY}
\end{align}
is the holomorphic function of $\tau$.
Also high order couplings are written by the products of $Y(\tau)$ \cite{Abe:2009dr}.
In what follows, we omit the chirality sign $\pm$ from the zero-modes.
The wavefunctions on $T^2_1 \times T^2_2 \times T^2_3$ are given by the products of the wavefunctions on $T^2_i$:
\begin{align}
  \psi_{T^2_1 \times T^2_2 \times T^2_3}^{j_1j_2j_3,|I_{ab}^{(1)}||I_{ab}^{(2)}||I_{ab}^{(3)}|}(z,\tau) = \psi^{j_1,|I_{ab}^{(1)}|}_{T^2_1}(z_1,\tau_1) \psi^{j_2,|I_{ab}^{(2)}|}_{T^2_2}(z_2,\tau_2) \psi^{j_3,|I_{ab}^{(3)}|}_{T^2_3}(z_3,\tau_3).
\end{align}
Thus, we have the  $|I_{ab}^{(1)}I_{ab}^{(2)}I_{ab}^{(3)}|$ generations of bi-fundamental zero-modes in our model construction.
Hereafter, we use the notation  $|I_{\alpha\beta}|$, $\alpha,\beta\in\{a,b,c\}$ as the magnetic flux.


\section{Modular symmetry on magnetized $T^2 \times T^2$}
\label{ModonT2xT2}

Here, we give a review on the modular symmetry of zero-modes.

\subsection{Modular symmetry and modular forms}
First, let us
review the modular symmetry on $T^2$ and the modular forms \cite{Gunning:1962,Schoeneberg:1974,Koblitz:1984,Bruinier:2008}.
As we mentioned at section \ref{U(N)}, the two-dimensional torus can be regarded as $T^2 \simeq \mathbb{C}/\Lambda$.
Here, $\Lambda$ is the two-dimensional lattice spanned by two basis vectors $e_k (k=1,2)$.
The complex coordinate on the torus, $z$, is defined by $z \equiv u/e_1$, where $u$ denotes the complex coordinate on $\mathbb{C}$.
The torus is characterized by the complex structure modulus $\tau$ defined by $\tau \equiv e_2/e_1 ({\rm Im}\tau>0)$.

Here, we can consider the following transformation of lattice vectors which span the same lattice,
\begin{align}
  \begin{pmatrix}
    e_2' \\
    e_1'
  \end{pmatrix}
  =
  \begin{pmatrix}
    a & b \\
    c & d
  \end{pmatrix}
  \begin{pmatrix}
    e_2 \\
    e_1
  \end{pmatrix}, \quad
  \gamma =
  \begin{pmatrix}
    a & b \\
    c & d
  \end{pmatrix}
  \in SL(2,\mathbb{Z}) \equiv \Gamma.
\end{align}
The $SL(2,\mathbb{Z}$) is generated by two generators:
\begin{align}
  S =
  \begin{pmatrix}
    0 & 1 \\
    -1 & 0
  \end{pmatrix}, \quad
  T =
  \begin{pmatrix}
    1 & 1 \\
    0 & 1
  \end{pmatrix},
\end{align}
satisfying the following algebraic relations,
\begin{align}
  S^2 \equiv Z = -\mathbb{I}, \quad S^4 = Z^2 = (ST)^3 = \mathbb{I}.
\end{align}
Under the $SL(2,\mathbb{Z})$ transformation, the complex coordinate on torus and the complex structure modulus are transformed as
\begin{align}
  &\gamma:z \equiv \frac{u}{e_1} \rightarrow z' \equiv \frac{u}{e_1'} = \frac{z}{c\tau + d}, \\
  &\gamma:\tau \equiv \frac{e_2}{e_1} \rightarrow \tau' \equiv \frac{e_2'}{e_1'} = \frac{a\tau + b}{c\tau + d}.
\end{align}
For $S$ and $T$ transformations, especially, they are transformed as
\begin{align}
  S: (z,\tau) \rightarrow \left(-\frac{z}{\tau},-\frac{1}{\tau}\right), \quad T: (z,\tau) \rightarrow (z,\tau + 1).\label{SandT}
\end{align}
Note that $\tau$ is invariant under $Z=-\mathbb{I}$, hence $\tau$ is transformed by $\bar{\Gamma}\equiv \Gamma/\{\pm\mathbb{I}\}$.

Now, we introduce the principal congruence subgroup of level $N$ defined as
\begin{align}
  \Gamma (N) \equiv \left\{ h=
  \begin{pmatrix}
    a' & b' \\
    c' & d'
  \end{pmatrix}
  \in \Gamma \left|
  \begin{pmatrix}
    a' & b' \\
    c' & d'
  \end{pmatrix}
  \equiv
  \begin{pmatrix}
    1 & 0 \\
    0 & 1
  \end{pmatrix}
  \right.
  ({\rm mod}\ N)
  \right\},
\end{align}
and its $\mathbb{Z}_2$ homomorphic group defined as $\bar{\Gamma}(N) \equiv \Gamma(N)/\{\pm\mathbb{I}\}$.
Then, we can define the quotient group $\Gamma_N \equiv \bar{\Gamma}/\bar{\Gamma}(N)$, 
where the following algebraic relations are satisfied \footnote{For $N\leq 5$, it is well known $\Gamma_2 \simeq S_3$, $\Gamma_3 \simeq A_4$, $\Gamma_4 \simeq S_4$ and $\Gamma_5 \simeq A_5$ \cite{deAdelhartToorop:2011re}. },
\begin{align}
S^2 = (ST)^3 = T^N = \mathbb{I} .
\end{align}
Similarly, the double covering group of $\Gamma_N$, $\Gamma_N'$,  is defined by $ \Gamma_N' \equiv \Gamma/\Gamma (N)$, 
where the following algebraic relations are satisfied,
\footnote{See e.g., Refs.~\cite{Liu:2019khw}.}
\begin{align}
  S^4=(ST)^3=T^N=\mathbb{I}, \qquad S^2T=TS^2 .
\end{align}

The modular forms, $f(\tau)$, of the integer weight\footnote{For $N=1,2$, the modular weight $k$ is only even because of the consistency of $-\mathbb{I} \in \Gamma(N)$.} $k$ for $\Gamma (N)$ are the holomorphic functions of $\tau$ transforming as
\begin{align}
  f(\gamma(\tau)) = J_k(\gamma,\tau) \rho(\gamma) f(\tau), \quad \gamma = 
  \begin{pmatrix}
    a & b \\
    c & d
  \end{pmatrix}
  \in \Gamma, \quad
  J_k(\gamma,\tau) = (c\tau + d)^k,
\end{align}
under $\Gamma$, where $\rho(\gamma)$ is the unitary representation of the quotient group $\Gamma_N'$ satisfying
\begin{align}
  \rho(\gamma_1\gamma_2) = \rho(\gamma_1)\rho(\gamma_2), \gamma_1,\gamma_2 \in \Gamma, \quad \rho(h) = \mathbb{I}, h \in \Gamma(N),
\end{align}
and the algebraic relations,
\begin{align}
  \rho(S)^4 = [\rho(S)\rho(T)]^3 = \rho(T)^N = \mathbb{I}, \quad \rho(S)^2\rho(T) = \rho(T)\rho(S)^2.
\end{align}
We note that since $\tau$ is invariant under $Z=-\mathbb{I}$, for even $k$, $\rho$ becomes a representation of $\Gamma_N$.

Next, we extend the modular forms to the half integer weight $k/2$, defining the double covering group of $\Gamma$, $\widetilde{\Gamma}$ as
\begin{align}
  \widetilde{\Gamma} &\equiv \{[\gamma,\epsilon]|\gamma\in\Gamma,\epsilon\in\{\pm 1\}\}
\end{align}
which is generated by two generators:
\begin{align}
  \widetilde{S} \equiv [S,1], \quad \widetilde{T} \equiv [T,1].
\end{align}
They satisfy the following algebraic relations:
\begin{align}
  \widetilde{S}^2=\widetilde{Z}, \qquad \widetilde{S}^4=(\widetilde{S}\widetilde{T})^3=\widetilde{Z}^2=[\mathbb{I},-1], 
\qquad \widetilde{S}^8=(\widetilde{S}\widetilde{T})^6=[\mathbb{I},1]\equiv\widetilde{\mathbb{I}}, 
\qquad \widetilde{Z}\widetilde{T} = \widetilde{T}\widetilde{Z} .
\end{align}
In the same way, the normal subgroup of $\widetilde{\Gamma}$, $\widetilde{\Gamma}(N)$, is defined by
\begin{align}
  \widetilde{\Gamma}(N) \equiv \{ [h,\epsilon]\in\widetilde{\Gamma} | h\in\Gamma(N),\epsilon = 1\}.
\end{align}
As the result, we can obtain the modular forms of half integer weight, $\widetilde{f}(\tau)$, as follows (See e.g., \cite{Koblitz:1984,shimura,Duncan:2018wbw}.):
\begin{align}
  \widetilde{f}(\widetilde{\gamma}(\tau))& = \widetilde{J}_{k/2} (\widetilde{\gamma},\tau) \widetilde{\rho}(\widetilde{\gamma}) \widetilde{f}(\tau), \quad \widetilde{\gamma} \in \widetilde{\Gamma}, \label{MFhalf} \\
  \widetilde{J}_{k/2}(\widetilde{\gamma},\tau) = &\epsilon^kJ_{k/2}(\gamma,\tau) = \epsilon^k (c\tau+d)^{k/2}, \quad k\in\mathbb{Z},
\end{align}
where $\widetilde{\rho}(\widetilde{\gamma})$ is the unitary representation of $\widetilde{\Gamma}_N \equiv \widetilde{\Gamma}/\widetilde{\Gamma}(N)$ satisfying
\begin{align}
  \widetilde{\rho}(\widetilde{\gamma}_1\widetilde{\gamma}_2) = \widetilde{\rho}(\widetilde{\gamma}_1)\widetilde{\rho}(\widetilde{\gamma}_2),\widetilde{\gamma}_1, \widetilde{\gamma}_2 \in \widetilde{\Gamma}, \quad  \widetilde{\rho}(\widetilde{h}) = \widetilde{\mathbb{I}}, \widetilde{h} \in \widetilde{\Gamma}(N),
\end{align}
and the algebraic relations,
\begin{align}
  \widetilde{\rho}(\widetilde{S})^2 = \widetilde{\rho}(\widetilde{Z}),\ \widetilde{\rho}(\widetilde{S})^4 = [\widetilde{\rho}(\widetilde{S})\widetilde{\rho}(\widetilde{T})]^3 = \widetilde{\rho}(\widetilde{Z})^2,\ \widetilde{\rho}(\widetilde{S})^8 = [\widetilde{\rho}(\widetilde{S})\widetilde{\rho}(\widetilde{T})]^6 = \widetilde{\mathbb{I}},\ \widetilde{\rho}(\widetilde{Z})\widetilde{\rho}(\widetilde{T}) = \widetilde{\rho}(\widetilde{T})\widetilde{\rho}(\widetilde{Z}).
\end{align}

\subsection{
Modular symmetry in magnetized $T^2$ models}
\label{The wavefunctions on magnetized T2}
Now, we are ready to express the wavefunctions on the magnetized torus as the terms of modular forms.
Under the modular transformation in Eq.~(\ref{SandT}), the zero-mode wavefunctions on the torus in Eq.~(\ref{psizero}) are transformed as
\begin{align}
  &S: \psi^{j,|I_{\alpha\beta}|}(z, \tau) \rightarrow \psi^{j,|I_{\alpha\beta}|}\left( -\frac{z}{\tau}, -\frac{1}{\tau} \right) = (-\tau)^{1/2} \sum_{k=0}^{|I_{\alpha\beta}|-1} e^{i\pi /4} \frac{1}{\sqrt{|I_{\alpha\beta}|}} e^{2\pi i \frac{jk}{|I_{\alpha\beta}|}} \psi^{k,|I_{\alpha\beta}|}(z, \tau), \label{psiS} \\
  &T: \psi^{j,|I_{\alpha\beta}|}(z, \tau) \rightarrow \psi^{j,|I_{\alpha\beta}|}(z, \tau+1) \ = e^{i\pi \frac{j^2}{|I_{\alpha\beta}|}} \psi^{j,|I_{\alpha\beta}|}(z, \tau). \label{psiT}
\end{align}
Note that this transformation can be considered only if the magnetic flux $|I_{\alpha\beta}|$ is even since the boundary conditions in Eq.~(\ref{BConTorus}) are variant under the $T$ transformation unless the flux is even.
Since the automorphy factor in Eq.~(\ref{psiS}), $(-\tau)^{1/2}$, means that the modular weight is $1/2$, we can apply
the modular forms of half integer weight in Eq.~(\ref{MFhalf}) to them as the following,
\begin{align}
  \psi^{j,|I_{\alpha\beta}|}(\widetilde{\gamma}(z,\tau)) &= \widetilde{J}_{1/2}(\widetilde{\gamma}, \tau) \sum_{k=0}^{|I_{\alpha\beta}|-1} \widetilde{\rho}(\widetilde{\gamma})_{jk} \psi^{k,|I_{\alpha\beta}|}(z,\tau), \quad \widetilde{\gamma} \in \widetilde{\Gamma}, \label{wavemodularform} \\
  \widetilde{\rho}(\widetilde{S})_{jk} =& e^{i\pi/4} \frac{1}{\sqrt{|I_{\alpha\beta}|}} e^{2\pi i\frac{jk}{|I_{\alpha\beta}|}}, \quad
\widetilde{\rho}(\widetilde{T})_{jk} = e^{i\pi \frac{j^2}{|I_{\alpha\beta}|}} \delta_{j,k}. \label{rhoSandT}
\end{align}
where $\widetilde{\rho}(\widetilde{\gamma})$ satisfies
\begin{align}
  \begin{array}{c}
    \widetilde{\rho}(\widetilde{S})^2 = \widetilde{\rho}(\widetilde{Z}),\ \widetilde{\rho}(\widetilde{S})^4 = [\widetilde{\rho}(\widetilde{S})\widetilde{\rho}(\widetilde{T})]^3 = -\widetilde{\mathbb{I}}, \ 
    \widetilde{\rho}(\widetilde{Z})\widetilde{\rho}(\widetilde{T}) = \widetilde{\rho}(\widetilde{T})\widetilde{\rho}(\widetilde{Z}),\ \widetilde{\rho}(\widetilde{T})^{2|I_{\alpha\beta}|} = \widetilde{\mathbb{I}}.
    \label{AlgebraonT2}
  \end{array}
\end{align}
Thus, the zero-mode wavefunctions on the magnetized $T^2$ can be regarded as the modular forms of weight $1/2$ for $\widetilde{\Gamma} (2|I_{\alpha\beta}|)$, and $\widetilde{\rho}(\widetilde{\gamma})$ is a unitary representation of the quotient group $\widetilde{\Gamma}_{2|I_{\alpha\beta}|} \equiv \widetilde{\Gamma}/\widetilde{\Gamma}(2|I_{\alpha\beta}|)$.

\subsection{
Modular symmetry in magnetized $T^2 \times T^2$ models}
Next, let us review the application of the modular forms to the zero-mode wavefunctions on the magnetized $T^2_1 \times T^2_2$.
Here and hereafter, we assume $T^2_1$ and $T^2_2$ have the same complex structure modulus, that is $\tau_1 = \tau_2 \equiv \tau$, 
because we will study the permutation between  $T^2_1$ and $T^2_2$ later.
Then, we can write down the modular transformation for the zero-modes on the $T^2_1 \times T^2_2$ as
\begin{align}
  \psi^{j,|I_{\alpha\beta}^{(1)}|}_{T^2_1}({\gamma}(z_1,\tau)) \psi^{k,|I_{\alpha\beta}^{(2)}|}_{T^2_2}({\gamma}(z_2,\tau))
  =& {J}_1({\gamma},\tau) \sum_{m=0}^{|I_{\alpha\beta}^{(1)}|-1}\widetilde{\rho}({\gamma})_{jm} \sum_{n=0}^{|I_{\alpha\beta}^{(2)}|-1}\widetilde{\rho}({\gamma})_{kn} 
  \psi^{m,|I_{\alpha\beta}^{(1)}|}_{T^2_1}(z_1,\tau) \psi^{n,|I_{\alpha\beta}^{(2)}|}_{T^2_2}(z_2,\tau) \notag \\
  \equiv& {J}_1({\gamma},\tau) \sum_{m=0}^{|I_{\alpha\beta}^{(1)}|-1} \sum_{n=0}^{|I_{\alpha\beta}^{(2)}|-1} \rho_{T^2_1 \times T^2_2}({\gamma})_{(jk)(mn)} \psi^{m,|I_{\alpha\beta}^{(1)}|}_{T^2_1}(z_1,\tau) \psi^{n,|I_{\alpha\beta}^{(2)}|}_{T^2_2}(z_2,\tau), 
\label{modularformT2xT2} \\
  \rho_{T^2_1 \times T^2_2}(S)_{(jk)(mn)} = \prod_{t=1,2} \widetilde{\rho}_{T^2_t}(\widetilde{S})_{j_tm_t} &
  = \widetilde{\rho}_{T^2_1}(\widetilde{S})_{jm} \widetilde{\rho}_{T^2_2}(\widetilde{S})_{kn}
  = \frac{i}{\sqrt{|I_{\alpha\beta}^{(1)}I_{\alpha\beta}^{(2)}|}} e^{2i\pi \left(\frac{jm}{|I_{\alpha\beta}^{(1)}|} + \frac{kn}{|I_{\alpha\beta}^{(2)}|}\right)},\\
  \rho_{T^2_1 \times T^2_2}(T)_{(jk)(mn)} = \prod_{t=1,2} \widetilde{\rho}_{T^2_t}(\widetilde{T})_{j_tm_t} &
  = \widetilde{\rho}_{T^2_1}(\widetilde{T})_{jm} \widetilde{\rho}_{T^2_2}(\widetilde{T})_{kn}
  = e^{i\pi \left(\frac{j^2}{|I_{\alpha\beta}^{(1)}|} + \frac{k^2}{|I_{\alpha\beta}^{(2)}|}\right)} \delta_{j,m} \delta_{k,n}, 
\label{SandTonT2xT2twist}\\
  j,m &\in \mathbb{Z}_{|I_{\alpha\beta}^{(1)}|},k,n \in \mathbb{Z}_{|I_{\alpha\beta}^{(2)}|}, {\gamma} \in {\Gamma}, \notag
\end{align}
where $\rho_{T^2_1 \times T^2_2}(\gamma)$  satisfies\footnote{lcm$(a,b)$ denotes the least common multiple of $a$ and $b$.}
\begin{align}
  \begin{array}{c}
    \rho(T)^{2{\rm lcm}(|I_{\alpha\beta}^{(1)}|,|I_{\alpha\beta}^{(2)}|)}_{(jk)(mn)} = \rho(S)^4_{(jk)(mn)} = [\rho(S)\rho(T)]^3_{(jk)(mn)} = \delta_{(jk),(mn)}, \\
    {[\rho(S)^2\rho(T)]}_{(jk)(mn)} = [\rho(T)\rho(S)^2]_{(jk)(mn)},\ \rho(S)^2_{(jk)(mn)} = -\delta_{j,|I_{\alpha\beta}^{(1)}|-m} \delta_{k,|I_{\alpha\beta}^{(2)}|-n}.
  \end{array}
\end{align}
Thus, the zero-mode wavefunctions on the magnetized $T^2_1 \times T^2_2$ can be regarded as the modular forms of weight $1$ for $\Gamma(2{\rm lcm}(|I_{\alpha\beta}^{(1)}|,|I_{\alpha\beta}^{(2)}|))$, 
and $\rho_{T^2_1 \times T^2_2}(\gamma)$ is a unitary representaion of the quotient group $\Gamma'_{2{\rm lcm}(|I_{\alpha\beta}^{(1)}|,|I_{\alpha\beta}^{(2)}|)}$.

In the following section, we consider several orbifold models and the modular symmetry.


\section{
Magnetized orbifold models}
\label{Magnetized orbifolds}
In this section, we study the zero-modes by orbifolding $T^2\times T^2$.
We also discuss the modular symmetry of the zero-modes
on the orbifolds.

\subsection{
Magnetized $T^2/\mathbb{Z}_2$ twisted orbifold models}
In this subsection, we briefly review the $T^2/\mathbb{Z}_2^{\rm(t)}$ twisted orbifold \cite{Abe:2008fi,Abe:2013bca,Abe:2014noa,Kobayashi:2017dyu}.
The $\mathbb{Z}_2^{\rm(t)}$ twist is defined by the following transformation of the complex coordinate on $T^2$:
\begin{align}
  z \rightarrow -z. \label{twist}
\end{align}
The zero-modes on $T^2$ are transformed as
\begin{align}
  \psi^{j,|I_{\alpha\beta}|}(z,\tau) \rightarrow \psi^{j,|I_{\alpha\beta}|}(-z,\tau) = \psi^{|I_{\alpha\beta}|-j,|I_{\alpha\beta}|}(z,\tau),
\end{align}
where the right hand side can be confirmed by the explict computation from Eq.~(\ref{psizero}).
The zero-modes on the $T^2/\mathbb{Z}_2^{\rm(t)}$ twisted orbifold, $\Psi^{j,|I_{\alpha\beta}|}_{T^2/\mathbb{Z}_2^{({\rm t})m}}(z)$, are required to obey the following boundary condition,
\begin{align}
  \Psi^{j,|I_{\alpha\beta}|}_{T^2/\mathbb{Z}_2^{{\rm(t)}m}}(e^{i\pi m}z) = e^{i\pi m} \Psi^{j,|I_{\alpha\beta}|}_{T^2/\mathbb{Z}_2^{{\rm(t)}m}}(z),
  \quad m \in \mathbb{Z}_2.
\end{align}
Thus, they can be expanded by the zero-modes on $T^2$ \cite{Abe:2008fi,Abe:2013bca,Abe:2014noa,Kobayashi:2017dyu} as
\begin{align}
  \Psi^{j,|I_{\alpha\beta}|}_{T^2/\mathbb{Z}_2^{{\rm(t)}m}}(z) = {\cal N}_{t,2}^j \left(\psi_{T^2}^{j,|I_{\alpha\beta}|}(z) + e^{i\pi m}\psi_{T^2}^{|I_{\alpha\beta}|-j,|I_{\alpha\beta}|}(z)\right),
  \quad {\cal N}_{t,2}^j = \left\{
  \begin{array}{l}
    1/2 \quad (j=0,|I_{\alpha\beta}|/2) \\
    1/\sqrt{2} \quad ({\rm otherwise})
  \end{array}
  \right.. \label{psitwist}
\end{align}
We find  the number of zero-modes on the $T^2/\mathbb{Z}_2^{\rm(t)}$ twisted orbifold as shown in Table \ref{NumZeroT}.
When  $|I_{\alpha\beta}| \in 2\mathbb{Z}$, the $\mathbb{Z}_2^{\rm(t)}$-even modes $(m = 0)$ and odd modes $(m = 1)$ 
have  $(|I_{\alpha\beta}|+2)/2$ and $(|I_{\alpha\beta}|-2)/2$ zero-modes, respectively.
On the other hand, when  $|I_{\alpha\beta}| \in 2\mathbb{Z}+1$, 
the $\mathbb{Z}_2^{\rm(t)}$-even modes $(m = 0)$ and odd modes $(m = 1)$ have 
$(|I_{\alpha\beta}|+1)/2$ and $ (|I_{\alpha\beta}|-1)/2$ zero-modes, respectively.
Then, we can obtain three generations from $\mathbb{Z}_2^{\rm(t)}$-even modes with $|I_{\alpha\beta}| = 4, 5$ and $\mathbb{Z}_2^{\rm(t)}$-odd modes with $|I_{\alpha\beta}| = 7, 8$.

\begin{table}[H]
  \begin{center}
  $\begin{array}{|c||ccccccccc|c|} \hline
    |I_{\alpha\beta}| & 0 & 2 & 4 & 6 & 8 & 10 & 12 & 14 & 16 & 2\mathbb{Z} \\ \hline
    {\rm even} & 1 & 2 & \fbox{3} & 4 & 5 & 6 & 7 & 8 & 9 & (|I_{\alpha\beta}|+2)/2 \\ \hline
    {\rm odd} & 0 & 0 & 1 & 2 & \fbox{3} & 4 & 5 & 6 & 7 & (|I_{\alpha\beta}|-2)/2 \\ \hline \hline
    |I_{\alpha\beta}| & 1 & 3 & 5 & 7 & 9 & 11 & 13 & 15 & 17 &  2\mathbb{Z}+1 \\ \hline
    {\rm even} & 1 & 2 & \fbox{3} & 4 & 5 & 6 & 7 & 8 & 9 & (|I_{\alpha\beta}|+1)/2 \\ \hline
    {\rm odd} & 0 & 1 & 2 & \fbox{3} & 4 & 5 & 6 & 7 & 8 & (|I_{\alpha\beta}|-1)/2 \\ \hline
  \end{array}$
  \caption{The number of zero-modes for $\mathbb{Z}_2^{\rm(t)}$-even and -odd on the $T^2/\mathbb{Z}_2^{\rm(t)}$ twisted orbifold with $|I_{\alpha\beta}| \leq 17$.
The three generations are boxed.}
  \label{NumZeroT}
  \end{center}
\end{table}

For $|I_{\alpha\beta}| \in 2\mathbb{Z}$, in particular, we can also study the modular symmetry of the zero-modes on the orbifold.
The zero-modes in Eq.~(\ref{psitwist}) transform as the modular forms of weight $1/2$ for $\widetilde{\Gamma}(2|I_{\alpha\beta}|)$ under the modular transformation:
\begin{align}
  \Psi^{j,|I_{\alpha\beta}|}_{T^2/\mathbb{Z}_2^{{\rm(t)}m}}(\widetilde{\gamma}(z,\tau)) = \widetilde{J}_{1/2}(\widetilde{\gamma},\tau) \sum_{k=0}^{|I_{\alpha\beta}|/2} \widetilde{\rho}_{T^2/\mathbb{Z}_2^{{\rm(t)}m}}(\widetilde{\gamma})_{jk} \Psi^{k,|I_{\alpha\beta}|}_{T^2/\mathbb{Z}_2^{{\rm(t)}m}}(z,\tau),
\end{align}
where
\begin{align}
  &\widetilde{\rho}_{T^2/\mathbb{Z}_2^{{\rm(t)}0}} (\widetilde{S})_{jk} = {\cal N}_{t,2}^j {\cal N}_{t,2}^k \frac{4e^{i\pi/4}}{\sqrt{|I_{\alpha\beta}|}} \cos \left(\frac{2\pi jk}{|I_{\alpha\beta}|}\right),\quad 
  \widetilde{\rho}_{T^2/\mathbb{Z}_2^{{\rm(t)}0}}(\widetilde{T})_{jk} = e^{i\pi \frac{j^2}{|I_{\alpha\beta}|}} \delta_{j,k}, \\
  &\widetilde{\rho}_{T^2/\mathbb{Z}_2^{{\rm(t)}1}} (\widetilde{S})_{jk} = {\cal N}_{t,2}^j {\cal N}_{t,2}^k \frac{4ie^{i\pi/4}}{\sqrt{|I_{\alpha\beta}|}} \sin \left(\frac{2\pi jk}{|I_{\alpha\beta}|}\right),\quad 
  \widetilde{\rho}_{T^2/\mathbb{Z}_2^{{\rm(t)}1}}(\widetilde{T})_{jk} = e^{i\pi \frac{j^2}{|I_{\alpha\beta}|}} \delta_{j,k}.
\end{align}
It is straightforward to check these unitary representations satisfy the algebraic relations in Eq.~(\ref{AlgebraonT2}).

\subsection{
Magnetized $(T^2_1 \times T^2_2)/\mathbb{Z}_2^{\rm(per)}$ permutation orbifold models}
In this subsection, we study the $\mathbb{Z}_2^{\rm(per)}$ permutation models of magnetized $(T^2_1 \times T^2_2)$, 
where the complex structure moduli are identified as $\tau_1 = \tau_2 \equiv \tau$.
The $\mathbb{Z}_2^{\rm(per)}$ permutation is defined by the following transformation of the complex coordinate on $T^2_1 \times T^2_2$:
\begin{align}
  (z_1,z_2) \rightarrow (z_2,z_1). \label{permutation}
\end{align}
The zero-modes on $T^2_1\times T^2_2$ are transformed under $\mathbb{Z}_2^{\rm(per)}$  as
\begin{align}
  \psi^{j,|I_{\alpha\beta}^{(1)}|}(z_1,\tau)\psi^{k,|I_{\alpha\beta}^{(2)}|}(z_2,\tau) \rightarrow \psi^{j,|I_{\alpha\beta}^{(1)}|}(z_2,\tau)\psi^{k,|I_{\alpha\beta}^{(2)}|}(z_1,\tau).
\end{align}
Since $T^2_1$ and $T^2_2$ cannot be distinguished because of the $\mathbb{Z}_2^{\rm(per)}$ permutation, we should also set the magnetic flux condition $|I_{\alpha\beta}^{(1)}| = |I_{\alpha\beta}^{(2)}| \equiv |I_{\alpha\beta}|$.
The zero-modes on the $(T^2_1 \times T^2_2)/\mathbb{Z}_2^{\rm(per)}$ permutation orbifold, $\Psi^{jk,|I_{\alpha\beta}|}_{(T^2_1 \times T^2_2)/\mathbb{Z}_2^{{\rm (per)}n}}(z_1,z_2)$, are required to obey the following boundary condition,
\begin{align}
  \Psi^{jk,|I_{\alpha\beta}|}_{(T^2_1 \times T^2_2)/\mathbb{Z}_2^{{\rm (per)}n}}(z_2,z_1) = e^{i\pi n} \Psi^{jk,|I_{\alpha\beta}|}_{(T^2_1 \times T^2_2)/\mathbb{Z}_2^{{\rm (per)}n}}(z_1,z_2),
  \quad n \in \mathbb{Z}_2.
\end{align}
Thus, they can be expanded by the zero-modes on $T^2$ \cite{Kikuchi:2020nxn} as
\begin{align}
  \Psi^{jk,|I_{\alpha\beta}|}_{(T^2_1 \times T^2_2)/\mathbb{Z}_2^{{\rm (per)}n}}(z_1,z_2) = {\cal N}_{2,p}^{jk} &\left(\psi^{j,|I_{\alpha\beta}|}(z_1)\psi^{k,|I_{\alpha\beta}|}(z_2)+e^{i\pi n}\psi^{j,|I_{\alpha\beta}|}(z_2)\psi^{k,|I_{\alpha\beta}|}(z_1)\right), \label{psipermutation} \\ 
  j,k\in\mathbb{Z}_{|I_{\alpha\beta}|}, &\quad j\geq k,\quad {\cal N}_{2,p}^{jk} = \left\{
  \begin{array}{l}
    1/2 \quad (j = k) \\
    1/\sqrt{2} \quad (j \neq k)
  \end{array}
  \right..\notag
\end{align}
Then, we find the number of zero-modes on the $(T^2_1 \times T^2_2)/\mathbb{Z}_2^{\rm (per)}$ permutation orbifold as shown in Table \ref{NumZeroP}.
The $\mathbb{Z}_2^{\rm(per)}$-even modes $(n = 0)$ and odd modes have $|I_{\alpha\beta}|(|I_{\alpha\beta}|+1)/2$  
and $|I_{\alpha\beta}|(|I_{\alpha\beta}|-1)/2$ zero-modes, respectively.
We can obtain three generations from $\mathbb{Z}_2^{\rm(per)}$-even modes with $|I_{\alpha\beta}| = 2$ and $\mathbb{Z}_2^{\rm(per)}$-odd modes with $|I_{\alpha\beta}| = 3$.

\begin{table}[H]
  \begin{center}
  $\begin{array}{|c||ccccccc|c|} \hline
    |I_{\alpha\beta}| & 0 & 1 & 2 & 3 & 4 & 5 & 6 & \mathbb{Z}  \\ \hline
    {\rm even} & 1 & 1 & \fbox{3} & 6 & 10 & 15 & 21 & |I_{\alpha\beta}|(|I_{\alpha\beta}|+1)/2 \\ \hline
    {\rm odd} & 0 & 0 & 1 & \fbox{3} & 6 & 10 & 15 & |I_{\alpha\beta}|(|I_{\alpha\beta}|-1)/2 \\ \hline
  \end{array}$
  \caption{The number of zero-modes for $\mathbb{Z}_2^{\rm(per)}$-even and -odd on the $(T^2_1 \times T^2_2)/\mathbb{Z}_2^{\rm(per)}$ permutation orbifold with $|I_{\alpha\beta}| \leq 6$.
The three generations are boxed.}
  \label{NumZeroP}
  \end{center}
\end{table}

For $|I_{\alpha\beta}| \in 2\mathbb{Z}$, in particular, we can also study the modular symmetry of the zero-modes on the orbifold.
The zero-modes in Eq.~(\ref{psipermutation}) transform as the modular forms of weight $1$ for $\Gamma (2|I_{\alpha\beta}|)$ under the modular transformation:
\begin{align}
  \Psi^{jk,|I_{\alpha\beta}|}_{(T^2_1 \times T^2_2)/\mathbb{Z}_2^{{\rm (per)}n}}(\gamma (z_1,z_2,\tau))
  = J_{1}(\gamma,\tau) \sum_{j'=0}^{|I_{\alpha\beta}|-1} \sum_{k'=0}^{j'}
  \rho_{(T^2_1\times T^2_2)/\mathbb{Z}_2^{{\rm (per)}n}} (\gamma)_{(jk)(j'k')}
  \Psi^{j'k',|I_{\alpha\beta}|}_{(T^2_1 \times T^2_2)/\mathbb{Z}_2^{{\rm (per)}n}}(z_1,z_2,\tau),
\end{align}
where
\begin{align}
  \rho_{(T^2_1\times T^2_2)/\mathbb{Z}_2^{{\rm (per)}n}} (\gamma)_{(jk)(j'k')}
  &= 2{\cal N}_{2,p}^{jk} {\cal N}_{2,p}^{j'k'} \left( \rho_{T^2_1\times T^2_2}(\gamma)_{(jk)(j'k')} + e^{i\pi n} \rho_{T^2_1\times T^2_2}(\gamma)_{(jk)(k'j')} \right), \label{UniReponPerm} \\
  &\rho_{T^2_1\times T^2_2}(\gamma)_{(jk)(j'k')} \equiv \widetilde{\rho}_{T^2_1}(\widetilde{\gamma})_{jj'} \widetilde{\rho}_{T^2_2}(\widetilde{\gamma})_{kk'}. \label{UniRepT2xT2}
\end{align}
In the following subsection, we will study the $(T^2_1\times T^2_2)/\mathbb{Z}_2^{\rm(per)}$ permutation orbifold 
combining the $\mathbb{Z}_2^{\rm(t)}$ twist.

\subsection{
Magnetized $(T^2_1 \times T^2_2)/(\mathbb{Z}_2^{\rm (t)}\times \mathbb{Z}_2^{\rm (per)})$  models }
In this subsection, we consider the $(T^2_1 \times T^2_2)/(\mathbb{Z}_2^{\rm (t)}\times \mathbb{Z}_2^{\rm (per)})$  models.
As in the previous subsection, we assume the same complex structure moduli $\tau_1 = \tau_2 \equiv \tau$ and magnetic flux $|I_{\alpha\beta}^{(1)}| = |I_{\alpha\beta}^{(2)}| \equiv |I_{\alpha\beta}|$ on $T^2_1 \times T^2_2$.
The wavefunctions on this orbifold, $\Psi^{jk,|I_{\alpha\beta}|}_{m({\rm t}),n({\rm per})}(z_1,z_2)$, 
are expanded by zero-modes on the $T^2/\mathbb{Z}_2^{\rm (t)}$ twisted orbifold, Eq.~(\ref{psitwist}),
as
\begin{align}
  \Psi^{jk,|I_{\alpha\beta}|}_{m({\rm t}),n({\rm per})}(z_1,z_2)
  = {\cal N}_{2,p}^{jk} &\left(\Psi^{j,|I_{\alpha\beta}|}_{T^2_1/\mathbb{Z}_2^{{\rm (t)}m}}(z_1)\Psi^{k,|I_{\alpha\beta}|}_{T^2_2/\mathbb{Z}_2^{{\rm (t)}m}}(z_2) + e^{i\pi n}\Psi^{j,|I_{\alpha\beta}|}_{T^2_2/\mathbb{Z}_2^{{\rm (t)}m}}(z_2)\Psi^{k,|I_{\alpha\beta}|}_{T^2_1/\mathbb{Z}_2^{{\rm (t)}m}}(z_1)\right), \label{psitp} \\
  j&,k \in \mathbb{Z}_{|I_{\alpha\beta}|/2+1},\quad j \geq k, \quad m,n \in \mathbb{Z}_2, \notag
\end{align}
Note that the $\mathbb{Z}_2^{\rm (t)}$ twist eigenvalue of the zero-modes on the $T^2_1/\mathbb{Z}_2^{\rm (t)}$ orbifold is the same as one on the $T^2_2/\mathbb{Z}_2^{\rm (t)}$ orbifold.
Then, we find  the number of zero-modes on the $(T^2_1 \times T^2_2)/(\mathbb{Z}_2^{\rm (t)}\times \mathbb{Z}_2^{\rm (per)})$  orbifold as shown in Table \ref{NumZeroTP}.
We can obtain three generations for ($\mathbb{Z}_2^{\rm (t)}$ twist, $\mathbb{Z}_2^{\rm (per)}$ permutation) = (even, even) modes with $|I_{\alpha\beta}| = 2, 3$, (even, odd) modes with $|I_{\alpha\beta}| = 4, 5$, (odd, even) modes with $|I_{\alpha\beta}| =5,  6$, and (odd, odd) modes with $|I_{\alpha\beta}| = 7, 8$.

\begin{table}[H]
  \begin{center}
  $\begin{array}{|c|ccccccccc|c|} \hline
    |I_{\alpha\beta}| & 0 & 2 & 4 & 6 & 8 & 10 & 12 & 14 & 16 &2\mathbb{Z}   \\ \hline
    {\rm (even,\ even)} & 1 & \fbox{3} & 6 & 10 & 15 & 21 & 28 & 36 & 45 & (|I_{\alpha\beta}|+2)(|I_{\alpha\beta}|+4)/8 \\ \hline
    {\rm (even,\ odd)} & 0 & 1 & \fbox{3} & 6 & 10 & 15 & 21 & 28 & 36 & |I_{\alpha\beta}|(|I_{\alpha\beta}|+2)/8 \\ \hline
    {\rm (odd,\ even)} & 0 & 0 & 1 & \fbox{3} & 6 & 10 & 15 & 21 & 28 & |I_{\alpha\beta}|(|I_{\alpha\beta}|-2)/8 \\ \hline
    {\rm (odd,\ odd)} & 0 & 0 & 0 & 1 & \fbox{3} & 6 & 10 & 15 & 21 & (|I_{\alpha\beta}|-2)(|I_{\alpha\beta}|-4)/8 \\ \hline \hline
    |I_{\alpha\beta}| & 1 & 3 & 5 & 7 & 9 & 11 & 13 & 15 & 17 & 2\mathbb{Z}+1 \\ \hline
    {\rm (even,\ even)} & 1 & \fbox{3} & 6 & 10 & 15 & 21 & 28 & 36 & 45 & (|I_{\alpha\beta}|+1)(|I_{\alpha\beta}|+3)/8 \\ \hline
    {\rm (even,\ odd)} & 0 & 1 & \fbox{3} & 6 & 10 & 15 & 21 & 28 & 36 & (|I_{\alpha\beta}|+1)(|I_{\alpha\beta}|-1)/8 \\ \hline
    {\rm (odd,\ even)} & 0 & 1 & \fbox{3} & 6 & 10 & 15 & 21 & 28 & 36 & (|I_{\alpha\beta}|+1)(|I_{\alpha\beta}|-1)/8 \\ \hline
    {\rm (odd,\ odd)} & 0 & 0 & 1 & \fbox{3} & 6 & 10 & 15 & 21 & 28 & (|I_{\alpha\beta}|-1)(|I_{\alpha\beta}|-3)/8 \\ \hline
  \end{array}$
  \caption{The numbers of zero-modes for ($\mathbb{Z}_2^{\rm (t)}$ twist, $\mathbb{Z}_2^{\rm (per)}$ permutation) = (even, even), (even, odd), (odd, even) and (odd, odd) on the $(T^2_1 \times T^2_2)/(\mathbb{Z}_2^{\rm (t)}\times \mathbb{Z}_2^{\rm (per)})$  orbifold with $|I_{\alpha\beta}| \leq 16$.
The three generations are boxed.}
  \label{NumZeroTP}
  \end{center}
\end{table}

For $|I_{\alpha\beta}| \in 2\mathbb{Z}$, in particular, we can also study the modular symmetry of the zero-modes on the orbifold.
The zero-modes in Eq.~(\ref{psitp}) transform as the modular forms of weight $1$ for $\Gamma (2|I_{\alpha\beta}|)$ under the modular transformation:
\begin{align}
  &\Psi^{jk,|I_{\alpha\beta}|}_{m({\rm t}),n({\rm per})}(\gamma (z_1,z_2,\tau))
  = J_{1}(\gamma,\tau) \sum_{j'=0}^{|I_{\alpha\beta}|/2} \sum_{k'=0}^{j'}
  \rho_{m({\rm t}),n({\rm per})} (\gamma)_{(jk)(j'k')}
  \Psi^{j'k',|I_{\alpha\beta}|}_{m({\rm t}),n({\rm per})}(z_1,z_2,\tau),
\end{align}
where
\begin{align}
  \rho_{m({\rm t}),n({\rm per})} (\gamma)_{(jk)(j'k')}
  &= 2{\cal N}_{2,p}^{jk} {\cal N}_{2,p}^{j'k'} \left( \rho_{(T^2_1\times T^2_2)/\mathbb{Z}_2^{\rm (t)}}(\gamma)_{(jk)(j'k')} + e^{i\pi n} \rho_{(T^2_1\times T^2_2)/\mathbb{Z}_2^{\rm (t)}}(\gamma)_{(jk)(k'j')} \right), \\
  &\rho_{(T^2_1\times T^2_2)/\mathbb{Z}_2^{\rm (t)}}(\gamma)_{(jk)(j'k')} \equiv \widetilde{\rho}_{T^2_1/\mathbb{Z}_2^{\rm (t)}}(\widetilde{\gamma})_{jj'} \widetilde{\rho}_{T^2_2/\mathbb{Z}_2^{\rm (t)}}(\widetilde{\gamma})_{kk'}.
\end{align}


\section{
Three generation models on magnetized $(T^2_1 \times T^2_2)/(\mathbb{Z}_2^{\rm (t)}\times \mathbb{Z}_2^{\rm (per)})$ orbifolds}
\label{ThreeonT2xT2}
In this section, we classfy all possible three generation models with non-vanishing Yukawa couplings.
In section \ref{U(N)}, we have seen the 10D $U(N)$ super-Yang-Mills theory on the background magnetic flux which breaks the gauge symmetry $U(N)$ into $U(N_a) \times U(N_b) \times U(N_c)$, where $N = N_a+N_b+N_c$.
Then, there are six types of bi-fundamental matter fields, $\lambda^{ab}(N_a,\bar{N}_b)$, $\lambda^{bc}(N_b,\bar{N}_c)$, $\lambda^{ca}(N_c,\bar{N}_a)$, $\lambda^{ba}(\bar{N}_a,N_b)$, $\lambda^{cb}(\bar{N}_b,N_c)$ and $\lambda^{ac}(\bar{N}_c,N_a)$.
When we focus on $ab$ sectors,
either $\lambda^{ab}(N_{a},\bar{N}_{b})$ or $\lambda^{ba}(\bar{N}_{a},N_{b})$ remains massless in the four-dimensional theory
after the chirality of the ten-dimensional gaugino fields is fixed.
It is also same for the other sectors.
In this section, we consider the gaugino $\lambda$ has the left (negative) chirality where only $\lambda^{ab}$, $\lambda^{bc}$ and $\lambda^{ca}$ appear.

For example, we can start with $U(8)$ gauge group and break to the Pati-Salam gauge group $SU(4) \times SU(2)_L \times SU(2)_R$ up to $U(1)$ factors, which contains the standard model gauge group, by choosing $N_a=4$, $N_b=2$ and $N_c=2$.
Note that some of the $U(1)$ part may be anomalous and massive by the Green-Schwarz mechanism.
Then, on our adoption of the chirality for gaugino fields, $\lambda^{ab}$ and $\lambda^{ca}$ correspond to left- and right-handed matter fields (quarks and leptons), respectively.
Similarly, $\lambda^{bc}$ is up-type and down-type Higgsinos.
Here, we assume supersymmetry is preserved at least locally at the $ab$, $bc$ and $ca$ sectors.
Thus, there are the same number of Higgs fields and Higgsino fields.
Furthermore, tachyonic modes do not exist at tree level.
Also we can break $U(8)$ to $SU(3) \times SU(2)_L \times (1)_Y$ up to $U(1)$ factors by proper combinations of 
magnetic fluxes and the $\mathbb{Z}_2^{\rm (t)}$ orbifold twist.
(See for details of model construction Refs.~\cite{Abe:2014vza,Abe:2017gye}.)
In these models, the three generations of left-handed quarks correspond to the gauginos $\lambda^{ab}$  originating from 
the open strings between $SU(3)$-brane and $SU(2)_L$ brane.
The up-sector (down-sector) of the right-handed quarks are originated from the open strings between 
the $SU(3)$-brane and $U(1)_c$ ($U(1)_{c'}$), i.e. $\lambda^{ca}$ ($\lambda^{c'a}$), 
where the up-sector and down-sector have different 
flavor structures.
The up-sector (down-sector) higgsinos are originated from 
the open strings between the $SU(2)$ and $U(1)_c$ ($U(1)_{c'}$), i.e. $\lambda^{bc}$ ($\lambda^{bc'}$).
Similarly, we can construct some realistic models from D7-brane models on $M^4\times T^2_1 \times T^2_2$.

As shown in section \ref{U(N)}, the extra dimensional part of the gaugino fields are written by the wavefunctions on the torus or its orbifolds.
Then, we can find three generation zero-modes on $T^2_1/\mathbb{Z}_2^{\rm(t)}$,
$(T^2_1 \times T^2_2)/\mathbb{Z}_2^{\rm (per)}$, and $(T^2_1 \times T^2_2)/(\mathbb{Z}_2^{\rm (t)} \times \mathbb{Z}_2^{\rm (per)})$ 
orbifolds by choosing the suitable fluxes as we have seen in section \ref{ModonT2xT2}.
If we realize three generations on the $T^2_1/\mathbb{Z}_2^{\rm(t)}$ orbifold,
a single zero-mode wavefunction must appear on the second and third tori, $T^2_1$ and $T^2_2$, to make the total generation three.
If we realize three generations on the $(T^2_1 \times T^2_2)/\mathbb{Z}_2^{\rm (per)}$ or 
$(T^2_1 \times T^2_2)/(\mathbb{Z}_2^{\rm (t)} \times \mathbb{Z}_2^{\rm (per)})$ orbifold,
a single zero-mode must appear on the third torus $T^2_3$.
In such  cases, the Yukawa matrix for 4D effective theory has rank three generally.
It is given by the overlap integral of zero-modes:
\begin{align}
  \widetilde{Y}^{IJK} = \int d^6z {\rm Tr}\left[\psi^I_L(z) \cdot \psi^J_R(z) \cdot \left(\phi^K_H(z)\right)^{\dagger}\right], \label{Overlap}
\end{align}
where $\psi_L^I$, $\psi_R^J$ and $\phi_H^K$ denote left-handed matter fields, right-handed matter fields and Higgs fields, respectively.
Note that the wavefunctions of bosons and fermions are the same.
In terms of three generation indices $I$ and $J$, Yukawa couplings can be written as
\begin{align}
  \widetilde{Y}^{IJ} = a^{IJ}_{(1\times 2)}(\tau)\cdot b_{(3)}(\tau_3),
\end{align}
where the lower indices $(1\times 2)$ and (3) mean the contributions from $T^2_1 \times T^2_2$ and $T^2_3$, respectively.
That is, $a_{(1 \times 2)}^{IJ}(\tau)$ is the function of $\tau$ originated from $T^2_1 \times T^2_2$ which determines the flavor structure of Yukawa couplings, and $b_{(3)}$ is the function of $\tau_3$ from $T^2_3$ which means overall factor.
However, the rank reduction of the matrix about $I$ and $J$ occurs in the case that $\psi_L^I$ and $\psi^J_R$ have the form such as
\begin{align}
  \psi^I_L = \psi^I_{T^2_1} \otimes \psi_{T^2_2}, \quad \psi^J_R = \psi_{T^2_1} \otimes \psi^J_{T^2_2},
\end{align}
where $\psi_{T^2_1}$ and $\psi_{T^2_2}$ are single zero-mode wavefunctions on $T^2_1$ and $T^2_2$, respectively.
It leads to the following form of Yukawa couplings,
\begin{align}
  a^{IJ}_{(1 \times 2)}(\tau) = {\alpha}^{I}_{(1)} \cdot {\beta}^{J}_{(2)},
\end{align}
whose rank is one.
We do not consider this type of flavor structure because it is not realistic phenomenologically.
Note that in the case that three generations come from only one torus, $T^2_1$, the wavefunctions are expressed as the following form:
\begin{align}
  \psi^I_L = \psi^I_{T^2_1} \otimes \psi_{T^2_2}, \quad \psi^J_R = \psi^J_{T^2_1} \otimes \psi_{T^2_2},
\end{align}
and the contributions from $T^2_2$ affect only on the overall factors, 
although the single zero-mode can affect to the modular symmetry of the wavefunctions.

Thus, we should consider the pair of three generation zero-modes
on the $T^2_1/\mathbb{Z}_2^{\rm(t)}$, 
$(T^2_1 \times T^2_2)/\mathbb{Z}_2^{\rm (per)}$, and $(T^2_1 \times T^2_2)/(\mathbb{Z}_2^{\rm (t)} \times \mathbb{Z}_2^{\rm (per)})$ 
 orbifolds, which coressponds to $\psi_L^I$ and $\psi_R^J$.
Then, 4D Yukawa couplings are required to be even under the $\mathbb{Z}_2^{\rm(t)}$ twist and $\mathbb{Z}_2^{\rm (per)}$ permutation. Othewise, it vanishes.
The mode patterns of $(\lambda^{ab},\lambda^{ca},\lambda^{bc})$ which lead to non-vanishing Yukawa couplings are shown in Table \ref{ModePatternTP}.

\begin{table}[H]
  \begin{center}
  $\begin{array}{c|ccc}
    & \lambda^{ab} & \lambda^{ca} & \lambda^{bc} \\ \hline
    {\rm I} & {\rm even} & {\rm even} & {\rm even} \\
    {\rm I\hspace{-.1em}I} & {\rm odd} & {\rm even} & {\rm odd} \\
    {\rm I\hspace{-.1em}I'} & {\rm even} & {\rm odd} & {\rm odd} \\
    {\rm I\hspace{-.1em}I\hspace{-.1em}I} & {\rm odd} & {\rm odd} & {\rm even} \\
  \end{array}$
  \caption{Possible $\mathbb{Z}_2$ twist (permutation) mode patterns of
zero-modes which lead to non-vanishing Yukawa couplings for one torus.}
  \label{ModePatternTP}
  \end{center}
\end{table}

\noindent
We note that the type ${\rm I\hspace{-.1em}I}$ is equivalent to type ${\rm I\hspace{-.1em}I'}$ by flipping left- and right-handed matter fields.
It means, for example, that the mode pattern $({\rm I\hspace{-.1em}I, I\hspace{-.1em}I})$ under ($\mathbb{Z}_2^{\rm (t)} $ twist, $\mathbb{Z}_2^{\rm (per)}$ permutation) is not equivalent to $({\rm I\hspace{-.1em}I, I\hspace{-.1em}I'})$ or $({\rm I\hspace{-.1em}I', I\hspace{-.1em}I})$ but equivalent to $({\rm I\hspace{-.1em}I', I\hspace{-.1em}I'})$.
Hence, we study either
obtained by left and right flipping.

Moreover, from the definition of the magnetic flux, the following relation should be satisfied:
\begin{align}
  |I_{cb}^{(i)}| = |I_{ab}^{(i)}| \pm |I_{ca}^{(i)}|, \quad i = 1,2,3. \label{FluxConditionThree}
\end{align}
It decides the magnetic flux for $\phi^K_H$ and also the number of Higgs fields.
Then, we can classify the three generation models with non-vanishing Yukawa couplings.
Results on the $T^2_1/\mathbb{Z}_2^{\rm (t)}$ orbifold were obtained in Ref.~\cite{Abe:2008sx},  which are 
shown in Tables \ref{ThreeGenT}.
Classifications of three generation models 
on $(T^2_1 \times T^2_2)/\mathbb{Z}_2^{\rm (per)}$, and $(T^2_1 \times T^2_2)/(\mathbb{Z}_2^{\rm (t)} \times \mathbb{Z}_2^{\rm (per)})$ 
orbifolds
are shown in Tables \ref{ThreeGenP} and  \ref{ThreeGenTP},  respectively.
In the following section, we study the explicit computation of Yukawa couplings and show ``the 2-2-4 model" on the 
$(T^2_1 \times T^2_2)/(\mathbb{Z}_2^{\rm (t)} \times \mathbb{Z}_2^{\rm (per)})$ orbifold as a simple example.

\begin{table}[H]
  \begin{center}
  \caption{Possible three generation models with non-vanishing Yukawa couplings on the $T^2_1/\mathbb{Z}_2^{\rm (t)} $ twisted orbifold.
The first column (T$_1$) shows the mode types shown in Table \ref{ModePatternTP} under $\mathbb{Z}_2^{\rm (t)} $ twist $({\rm T_1})$.
The second column shows the magnetic fluxes which give three generations for left ($ab$) and right ($ca$) handed matter fields.
The symbol ``/'' distinguishes the choise of sign $\pm$ on the magnetix fluxes condition in Eq.~(\ref{FluxConditionThree}).
The remaining columns show the algebraic relations for the modular transformations $\widetilde{\rho}^{(\alpha\beta)}$ and $\widetilde{\rho}^{(Y)}$ acting on the $\alpha\beta$-sector wavefunctions, $\alpha\beta \in \{ab,ca,cb\}$, and the Yukawa couplings, respectively.
\newline}
  \label{ThreeGenT}
  \renewcommand{\arraystretch}{1.2}
  $\begin{array}{c|ccc|c|c|c|c|c} \hline
    \multicolumn{9}{c}{T^2_1/\mathbb{Z}_2^{\rm (t)} \ {\rm twisted\ orbifold}} \\ \hline \hline
    ({\rm T_1}) & |I_{ab}^{(1)}| & |I_{ca}^{(1)}| & |I_{cb}^{(1)}| & \begin{array}{c}{\rm The\ number\ of} \\ {\rm Higgs\ zero}$-${\rm modes}\end{array} & \widetilde{\rho}^{(ab)} & \widetilde{\rho}^{(ca)} & \widetilde{\rho}^{(cb)} & \widetilde{\rho}^{(Y)} \\ \hline\hline
    ({\rm I}) & 4 & 4 & 8/0 & 5/1 & \widetilde{\Gamma}'_{8} & \widetilde{\Gamma}'_{8} & \widetilde{\Gamma}'_{16}/$-$ & \widetilde{\Gamma}'_{16}/$-$ \\
    ({\rm I\hspace{-.1em}I}) & 8 & 4 & 12/4 & 5/1 & \widetilde{\Gamma}'_{16} & \widetilde{\Gamma}'_{8} & \widetilde{\Gamma}'_{24}/\widetilde{\Gamma}'_{8} & \widetilde{\Gamma}'_{48}/\widetilde{\Gamma}'_{16} \\
    ({\rm I\hspace{-.1em}I\hspace{-.1em}I}) & 8 & 8 & 16/0 & 9/1 & \widetilde{\Gamma}'_{16} & \widetilde{\Gamma}'_{16} & \widetilde{\Gamma}'_{32}/$-$ & \widetilde{\Gamma}'_{32}/$-$ \\
    ({\rm I}) & 5 & 4 & 9/1 & 5/1 & $-$ & \widetilde{\Gamma}'_{8} & $-$ & \widetilde{\Gamma}'_{360}/\widetilde{\Gamma}'_{40} \\
    ({\rm I}) & 5 & 5 & 10/0 & 6/1 & $-$ & $-$ & \widetilde{\Gamma}'_{20}/$-$ & \widetilde{\Gamma}'_{500}/$-$ \\
    ({\rm I\hspace{-.1em}I}) & 8 & 5 & 13/3 & 6/1 & \widetilde{\Gamma}'_{16} & $-$ & $-$ & \widetilde{\Gamma}'_{1040}/\widetilde{\Gamma}'_{240} \\
    ({\rm I\hspace{-.1em}I}) & 7 & 4 & 11/3 & 5/1 & $-$ & \widetilde{\Gamma}'_{8} & $-$ & \widetilde{\Gamma}'_{616}/\widetilde{\Gamma}'_{168} \\
    ({\rm I\hspace{-.1em}I}) & 7 & 5 & 12 & 5 & $-$ & $-$ & \widetilde{\Gamma}'_{24} & \widetilde{\Gamma}'_{840} \\
    ({\rm I\hspace{-.1em}I\hspace{-.1em}I}) & 8 & 7 & 15/1 & 8/1 & \widetilde{\Gamma}'_{16} & $-$ & $-$ & \widetilde{\Gamma}'_{1680}/\widetilde{\Gamma}'_{112} \\
    ({\rm I\hspace{-.1em}I\hspace{-.1em}I}) & 7 & 7 & 14/0 & 8/1 & $-$ & $-$ & \widetilde{\Gamma}'_{28}/$-$ & \widetilde{\Gamma}'_{1372}/$-$ \\ \hline
  \end{array}$
  \end{center}
\end{table}

\begin{table}[H]
  \begin{center}
  \caption{Possible three generation models with non-vanishing Yukawa couplings on the $(T^2_1 \times T^2_2)/\mathbb{Z}_2^{\rm (per)} $ permutation orbifold.
The first column (P) means the mode types shown in Table \ref{ModePatternTP} under $\mathbb{Z}_2^{\rm (per)}$ permutation (P).
Note that the magnetic fluxes for $T^2_1$ are the same as ones for $T^2_2$ because of the $\mathbb{Z}_2^{\rm (per)}$ permutation, 
that is, $|I_{\alpha\beta}^{(1)}|=|I_{\alpha\beta}^{(2)}|\equiv |I_{\alpha\beta}|$, $\alpha\beta \in \{ab,ca,cb\}$.
\newline}
  \label{ThreeGenP}
  \renewcommand{\arraystretch}{1.2}
  $\begin{array}{c|ccc|c|c|c|c|c} \hline
    \multicolumn{9}{c}{(T^2_1\times T^2_2)/\mathbb{Z}_2^{\rm (per)}\ {\rm permutation\ orbifold}} \\ \hline \hline
    {\rm (P)} & |I_{ab}| & |I_{ca}| & |I_{cb}| & \begin{array}{c}{\rm The\ number\ of}\\ {\rm Higgs\ zero}$-${\rm modes}\end{array} & \rho^{(ab)} & \rho^{(ca)} & \rho^{(cb)} & \rho^{(Y)} \\ \hline \hline
    ({\rm I}) & 2 & 2 & 4/0 & 10/1 & {\Gamma}'_{4} & {\Gamma}'_{4} & {\Gamma}'_{8}/$-$ & {\Gamma}'_{8}/$-$ \\
    ({\rm I\hspace{-.1em}I}) & 3 & 2 & 5 & 10 & $-$ & {\Gamma}'_{4} & $-$ & {\Gamma}'_{60} \\
    ({\rm I\hspace{-.1em}I\hspace{-.1em}I}) & 3 & 3 & 6/0 & 21/1 & $-$ & $-$ & {\Gamma}'_{12}/$-$ & {\Gamma}'_{108}/$-$ \\ \hline
  \end{array}$
  \end{center}
\end{table}

\begin{table}[H]
  \begin{center}
  \caption{Possible three generation models with non-vanishing Yukawa couplings on the 
$(T^2_1 \times T^2_2)/(\mathbb{Z}_2^{\rm (t)} \times \mathbb{Z}_2^{\rm (per)})$ orbifold.
The first column (T,P) shows the mode types shown in Table \ref{ModePatternTP} under $\mathbb{Z}_2^{\rm (t)} $ twist (T) and $\mathbb{Z}_2^{\rm (per)})$ permutation (P). 
\newline}
  \label{ThreeGenTP}
  \renewcommand{\arraystretch}{1.2}
  $\begin{array}{c|ccc|c|c|c|c|c} \hline
    \multicolumn{9}{c}{(T^2_1\times T^2_2)/(\mathbb{Z}_2^{\rm (t)} \times \mathbb{Z}_2^{\rm (per)})\ {\rm orbifold}} \\ \hline \hline
    {\rm (T,P)} & |I_{ab}| & |I_{ca}| & |I_{cb}| & \begin{array}{c}{\rm The\ number\ of}\\ {\rm Higgs\ zero}$-${\rm modes}\end{array} & \rho^{(ab)} & \rho^{(ca)} & \rho^{(cb)} & \rho^{(Y)} \\ \hline \hline
    ({\rm I},{\rm I}) & 2 & 2 & 4/0 & 6/1 & {\Gamma}'_{4} & {\Gamma}'_{4} & {\Gamma}'_{8}/$-$ & {\Gamma}'_{8}/$-$ \\
    ({\rm I},{\rm I\hspace{-.1em}I}) & 4 & 2 & 6/2 & 6/1 & {\Gamma}'_{8} & {\Gamma}'_{4} & {\Gamma}'_{12}/{\Gamma}'_{4} & {\Gamma}'_{24}/{\Gamma}'_{8} \\
    ({\rm I},{\rm I\hspace{-.1em}I\hspace{-.1em}I}) & 4 & 4 & 8/0 & 15/1 & {\Gamma}'_{8} & {\Gamma}'_{8} & {\Gamma}'_{16}/$-$ & {\Gamma}'_{16}/$-$ \\
    ({\rm I\hspace{-.1em}I},{\rm I}) & 6 & 2 & 8/4 & 6/1 & {\Gamma}'_{12} & {\Gamma}'_{4} & {\Gamma}'_{16}/{\Gamma}'_{8} & {\Gamma}'_{48}/{\Gamma}'_{24} \\
    ({\rm I\hspace{-.1em}I},{\rm I\hspace{-.1em}I}) & 8 & 2 & 10/6 & 6/1 & {\Gamma}'_{16} & {\Gamma}'_{4} & {\Gamma}'_{20}/{\Gamma}'_{12} & {\Gamma}'_{80}/{\Gamma}'_{48} \\
    ({\rm I\hspace{-.1em}I},{\rm I\hspace{-.1em}I'}) & 6 & 4 & 10 & 6 & {\Gamma}'_{12} & {\Gamma}'_{8} & {\Gamma}'_{20} & {\Gamma}'_{120} \\
    ({\rm I\hspace{-.1em}I},{\rm I\hspace{-.1em}I\hspace{-.1em}I}) & 8 & 4 & 12/4 & 15/1 & {\Gamma}'_{16} & {\Gamma}'_{8} & {\Gamma}'_{24}/{\Gamma}'_{8} & {\Gamma}'_{48}/{\Gamma}'_{16} \\
    ({\rm I\hspace{-.1em}I\hspace{-.1em}I},{\rm I}) & 6 & 6 & 12/0 & 28/1 & {\Gamma}'_{12} & {\Gamma}'_{12} & {\Gamma}'_{24}/$-$ & {\Gamma}'_{24}/$-$ \\
    ({\rm I\hspace{-.1em}I\hspace{-.1em}I},{\rm I\hspace{-.1em}I}) & 8 & 6 & 14/2 & 28/1 & {\Gamma}'_{16} & {\Gamma}'_{12} & {\Gamma}'_{28}/{\Gamma}'_{4} & {\Gamma}'_{336}/{\Gamma}'_{48} \\
    ({\rm I\hspace{-.1em}I\hspace{-.1em}I},{\rm I\hspace{-.1em}I\hspace{-.1em}I}) & 8 & 8 & 16/0 & 45/1 & {\Gamma}'_{16} & {\Gamma}'_{16} & {\Gamma}'_{32}/$-$ & {\Gamma}'_{32}/$-$ \\
 \hline
  \end{array}$
  \end{center}
\end{table}

\begin{table}[H]
  \begin{center}
  \renewcommand{\arraystretch}{1.2}
  $\begin{array}{c|ccc|c|c|c|c|c} \hline
    \multicolumn{9}{c}{(T^2_1\times T^2_2)/(\mathbb{Z}_2^{\rm (t)} \times \mathbb{Z}_2^{\rm (per)})\ {\rm orbifold}} \\ \hline \hline
    {\rm (T,P)} & |I_{ab}| & |I_{ca}| & |I_{cb}| & \begin{array}{c}{\rm The\ number\ of}\\ {\rm Higgs\ zero}$-${\rm modes}\end{array} & \rho^{(ab)} & \rho^{(ca)} & \rho^{(cb)} & \rho^{(Y)} \\ \hline \hline
    ({\rm I},{\rm I}) & 3 & 2 & 5/1 & 6/1 & $-$ & {\Gamma}'_{4} & $-$ & {\Gamma}'_{60}/{\Gamma}'_{12} \\
    ({\rm I},{\rm I}) & 3 & 3 & 6/0 & 10/1 & $-$ & $-$ & {\Gamma}'_{12}/$-$ & {\Gamma}'_{108}/$-$ \\
    ({\rm I},{\rm I\hspace{-.1em}I}) & 4 & 3 & 7 & 6 & {\Gamma}'_{8} & $-$ & $-$ & {\Gamma}'_{168} \\
    ({\rm I},{\rm I\hspace{-.1em}I}) & 5 & 2 & 7/3 & 6/1 & $-$ & {\Gamma}'_{4} & $-$ & {\Gamma}'_{140}/{\Gamma}'_{60} \\
    ({\rm I},{\rm I\hspace{-.1em}I}) & 5 & 3 & 8/2 & 10/1 & $-$ & $-$ & {\Gamma}'_{16}/{\Gamma}'_{4} & {\Gamma}'_{240}/{\Gamma}'_{60} \\
    ({\rm I},{\rm I\hspace{-.1em}I\hspace{-.1em}I}) & 5 & 4 & 9/1 & 15/1 & $-$ & {\Gamma}'_{8} & $-$ & {\Gamma}'_{360}/{\Gamma}'_{40} \\
    ({\rm I},{\rm I\hspace{-.1em}I\hspace{-.1em}I}) & 5 & 5 & 10/0 & 21/1 & $-$ & $-$ & {\Gamma}'_{20}/$-$ & {\Gamma}'_{500}/$-$ \\
    ({\rm I\hspace{-.1em}I},{\rm I}) & 6 & 3 & 9/3 & 10/1 & {\Gamma}'_{12} & $-$ & $-$ & {\Gamma}'_{324}/{\Gamma}'_{108} \\
    ({\rm I\hspace{-.1em}I},{\rm I}) & 5 & 2 & 7/3 & 6/1 & $-$ & {\Gamma}'_{4} & $-$ & {\Gamma}'_{140}/{\Gamma}'_{60} \\
    ({\rm I\hspace{-.1em}I},{\rm I}) & 5 & 3 & 8 & 6 & $-$ & $-$ & {\Gamma}'_{16} & {\Gamma}'_{240} \\
    ({\rm I\hspace{-.1em}I},{\rm I\hspace{-.1em}I}) & 8 & 3 & 11/5 & 10/1 & {\Gamma}'_{16} & $-$ & $-$ & {\Gamma}'_{528}/{\Gamma}'_{240} \\
    ({\rm I\hspace{-.1em}I},{\rm I\hspace{-.1em}I}) & 7 & 2 & 9/5 & 6/1 & $-$ & {\Gamma}'_{4} & $-$ & {\Gamma}'_{252}/{\Gamma}'_{140} \\
    ({\rm I\hspace{-.1em}I},{\rm I\hspace{-.1em}I}) & 7 & 3 & 10 & 6 & $-$ & $-$ & {\Gamma}'_{20} & {\Gamma}'_{420} \\
    ({\rm I\hspace{-.1em}I},{\rm I\hspace{-.1em}I'}) & 6 & 5 & 11 & 10 & {\Gamma}'_{12} & $-$ & $-$ & {\Gamma}'_{660} \\
    ({\rm I\hspace{-.1em}I},{\rm I\hspace{-.1em}I'}) & 5 & 4 & 9 & 6 & $-$ & {\Gamma}'_{8} & $-$ & {\Gamma}'_{360} \\
    ({\rm I\hspace{-.1em}I},{\rm I\hspace{-.1em}I'}) & 5 & 5 & 10 & 6 & $-$ & $-$ & {\Gamma}'_{20} & {\Gamma}'_{500} \\
    ({\rm I\hspace{-.1em}I},{\rm I\hspace{-.1em}I\hspace{-.1em}I}) & 8 & 5 & 13/3 & 21/1 & {\Gamma}'_{16} & $-$ & $-$ & {\Gamma}'_{1040}/{\Gamma}'_{240} \\
    ({\rm I\hspace{-.1em}I},{\rm I\hspace{-.1em}I\hspace{-.1em}I}) & 7 & 4 & 11/3 & 15/1 & $-$ & {\Gamma}'_{8} & $-$ & {\Gamma}'_{616}/{\Gamma}'_{168} \\
    ({\rm I\hspace{-.1em}I},{\rm I\hspace{-.1em}I\hspace{-.1em}I}) & 7 & 5 & 12 & 15 & $-$ & $-$ & {\Gamma}'_{24} & {\Gamma}'_{840}/{\Gamma}'_{140} \\
    ({\rm I\hspace{-.1em}I\hspace{-.1em}I},{\rm I}) & 6 & 5 & 11/1 & 21/1 & {\Gamma}'_{12} & $-$ & $-$ & {\Gamma}'_{660}/{\Gamma}'_{60} \\
    ({\rm I\hspace{-.1em}I\hspace{-.1em}I},{\rm I}) & 5 & 5 & 10/0 & 21/1 & $-$ & $-$ & {\Gamma}'_{20}/$-$ & {\Gamma}'_{500}/$-$ \\
    ({\rm I\hspace{-.1em}I\hspace{-.1em}I},{\rm I\hspace{-.1em}I}) & 8 & 5 & 13/3 & 21/1 & {\Gamma}'_{16} & $-$ & $-$ & {\Gamma}'_{1040}/{\Gamma}'_{240} \\
    ({\rm I\hspace{-.1em}I\hspace{-.1em}I},{\rm I\hspace{-.1em}I}) & 7 & 6 & 13 & 21 & $-$ & {\Gamma}'_{12} & $-$ & {\Gamma}'_{1092} \\
    ({\rm I\hspace{-.1em}I\hspace{-.1em}I},{\rm I\hspace{-.1em}I}) & 7 & 5 & 12/2 & 21/1 & $-$ & $-$ & {\Gamma}'_{24}/{\Gamma}'_{4} & {\Gamma}'_{840}/{\Gamma}'_{140} \\
    ({\rm I\hspace{-.1em}I\hspace{-.1em}I},{\rm I\hspace{-.1em}I\hspace{-.1em}I}) & 8 & 7 & 15/1 & 36/1 & {\Gamma}'_{16} & $-$ & $-$ & {\Gamma}'_{1680}/{\Gamma}'_{112} \\
    ({\rm I\hspace{-.1em}I\hspace{-.1em}I},{\rm I\hspace{-.1em}I\hspace{-.1em}I}) & 7 & 7 & 14/0 & 36/1 & $-$ & $-$ & {\Gamma}'_{28}/$-$ & {\Gamma}'_{1372}/$-$ \\ \hline
  \end{array}$
  \end{center}
\end{table}


\section{Yukawa couplings in three generation models}
\label{Yukawa}


\subsection{Yukawa interactions}
\label{YukawaInt}

As mentioned at section \ref{ThreeonT2xT2}, Yukawa coupling for 4D effective theory is given by the overlap integral of zero-modes
in Eq.~(\ref{Overlap}) as
\begin{align}
  \widetilde{Y}^{ijk} = \int_{T^2} dz d\bar{z} \psi_{T^2}^{i,|I_{ab}|}(z) \cdot \psi_{T^2}^{j,|I_{ca}|}(z) \cdot \left(\psi_{T^2}^{k,|I_{cb}|}(z)\right)^*.
  \label{YukawaonT2}
\end{align}
It can be calculated by using the normalization in Eq.~(\ref{Normalization}) and the product expansions in Eq.(\ref{ProductExpansion}) as the following:
\begin{align}
  \widetilde{Y}^{ijk} = (2{\rm Im}\tau)^{-1/2} {\cal A}^{-1/2} \left|\frac{I_{ab}I_{ca}}{I_{cb}}\right|^{1/4} \sum_{m=0}^{|I_{cb}|-1} \vartheta
  \begin{bmatrix}
    \frac{|I_{ca}|i-|I_{ab}|j+|I_{ab}I_{ca}|m}{|I_{ab}I_{ca}I_{cb}|} \\ 0
  \end{bmatrix}
  (0, |I_{ab}I_{ca}I_{cb}|\tau) \times \delta_{i+j-k,|I_{cb}|\ell-|I_{ab}|m},
\end{align}
where $\ell \in \mathbb{Z}$.
There are the selection rule $i+j-k \propto g \equiv {\rm gcd}(|I_{ab}|,|I_{cb}|)$ for nonzero Yukawa couplings\footnote{gcd$(a,b)$ denotes the greatest common divisor of $a$ and $b$.}.
If we find a combination of indices $i,j,k$ satisfying this selection rule, there is one integer solution such as $|m| \in \mathbb{Z}_{|I_{cb}|/g}$, $|\ell| \in \mathbb{Z}_{|I_{ab}|/g}$.
Thus, we can rewrite it as the sum of $g$ number of theta functions:
\begin{align}
  \widetilde{Y}^{ijk} = (2{\rm Im}\tau)^{-1/2} {\cal A}^{-1/2} \left|\frac{I_{ab}I_{ca}}{I_{cb}}\right|^{1/4} \sum_{n=1}^{g} \vartheta
  \begin{bmatrix}
    \frac{|I_{ca}|k-|I_{cb}|j+|I_{ca}I_{cb}|\ell_0}{|I_{ab}I_{ca}I_{cb}|} + \frac{n}{g} \\ 0
  \end{bmatrix}
  (0, |I_{ab}I_{ca}I_{cb}|\tau) \label{Ytilde}
\end{align}
where $\ell_0$ is the integer satisfying $i+j-k=|I_{cb}|\ell_0-|I_{ab}|m_0$, $m_0 \in \mathbb{Z}$.
Hereafter, we use the following notation:
\begin{align}
  \eta_N(\tau) = \vartheta
  \begin{bmatrix}
    \frac{N}{M} \\
    0
  \end{bmatrix}
  (0,M\tau), \quad M = I_{ab}I_{ca}I_{cb}. \label{etanotation}
\end{align}
When we calculate the Yukawa couplings on the orbifolds, we should calculate the overlap integral of zero-modes
on the orbifolds, and hence take the proper liner combinations of the zero-modes
on the torus as we have seen in section \ref{Magnetized orbifolds}.

We can also extend this analysis to the
Yukawa couplings on $T^2_1 \times T^2_2$ and orbifolds using the
zero-modes on each space; for example, Yukawa couplings on $T^2_1 \times T^2_2$ are given by
\begin{align}
  \widetilde{Y}^{IJK} &= \int_{T^2_1 \times T^2_2} dz_1^2 dz_2^2 \left(\psi_{T^2_1}^{i^{(1)},|I_{ab}^{(1)}|}\psi_{T^2_2}^{i^{(2)},|I_{ab}^{(2)}|}\right) \cdot \left(\psi_{T^2_1}^{j^{(1)},|I_{ca}^{(1)}|}\psi_{T^2_2}^{j^{(2)},|I_{ca}^{(2)}|}\right) \cdot \left(\psi_{T^2_1}^{k^{(1)},|I_{cb}^{(1)}|}\psi_{T^2_2}^{k^{(2)},|I_{cb}^{(2)}|}\right)^* \notag \\
  &= \prod_{r=1,2}\left[\int_{T^2_r} dz_r d\bar{z}_r \psi_{T^2_r}^{i^{(r)},|I_{ab}^{(r)}|}(z_r) \cdot \psi_{T^2_r}^{j^{(r)},|I_{ca}^{(r)}|}(z_r) \cdot \left(\psi_{T^2_r}^{k^{(r)},|I_{cb}^{(r)}|}(z_r)\right)^*\right] \notag \\
  &= \prod_{r=1,2} \widetilde{Y}^{ijk}_{(r)}
\end{align}
where $I=(i_1i_2), J=(j_1j_2), K=(k_1k_2)$.

In the previous section, we have summarized the possible three generation models with non-vanishing Yukawa couplings.
There are several models with $|I_{cb}|=0$,
where Higgs zero-mode
is contant.
In such cases, by using the
normalization in Eq.~(\ref{Normalization}) and the fluxes relation in Eq.~(\ref{FluxConditionThree}), it can be checked that the Yukawa coupling is proportional to the $(3 \times 3)$ unit matrix.
This is not realistic and should not be considered.


\subsection{Modular symmetry of Yukawa couplings}
\label{ModforYukawa}
Here, we study the modualr symmetry in the Yukawa couplings.
Since Yukawa coupling on $T^2$ is the overlap integration of products of three zero-modes,
we can express it by the terms of the modular forms as shown in section \ref{The wavefunctions on magnetized T2}.

The superpotential in global supersymmetric theory, $\widetilde{W}$, is related to one in the supergravity theory, $W$ as
\begin{align}
|\widetilde{W}|^2 = e^{K} |W|^2,
\end{align}
where 
$K$ is the K\"{a}hler potential and is written for the modulus $\tau$ as 
\begin{align}
K = - \ln{[i(\bar{\tau}-\tau)]} = - \ln{(2{\rm Im}\tau)}.
\end{align}
Indeed, the Yukawa coupling in the global supersymmetric model (\ref{Ytilde}) can be written as
\begin{align}
\widetilde{Y}^{ijk} = e^{K/2} Y^{ijk}(\tau), 
\end{align}
where $Y^{ijk}(\tau)$ is the holomorphic function of $\tau$ in Eq.~(\ref{productY}), i.e.,
\begin{align}
Y^{ijk}(\tau) = {\cal A}^{-1/2} \left|\frac{I_{ab}I_{ca}}{I_{cb}}\right|^{1/4} \sum_{n=1}^{g} \eta_{\left(|I_{ca}|k-|I_{cb}|j+|I_{ca}I_{cb}|\ell_0+n\frac{|I_{ab}I_{ca}I_{cb}|}{g}\right)}(\tau). \label{holomY}
\end{align}
Thus, the Yukawa coupling  $Y^{ijk}(\tau)$ corresponds  to the holomorphic Yukawa couplings in supergravity basis.
Hereafter, we study the modular symmetry of the holomorphic Yukawa couplings $Y^{ijk}(\tau)$ in Eq.~(\ref{holomY}) and find that they transform as the modular forms.

First, we can show
\begin{align}
  \left( |I_{ca}|k-|I_{cb}|j+|I_{ca}I_{cb}|\ell_{0} + n|I_{ab}I_{ca}I_{cb}|/g \right) \in \mathbb{Z}_{|I_{ab}I_{ca}I_{cb}|} \quad {\rm for}\ j\in\mathbb{Z}_{|I_{ca}|},\ k \in \mathbb{Z}_{|I_{cb}|}, \ n \in \mathbb{Z}_g,
\end{align}
and rewrite
Eq.~(\ref{holomY}) as
\begin{align}
  Y^{ijk}(\tau) = c\sum_{n=1}^g \eta_{g\left(J+(n-1)\frac{|I_{ab}I_{ca}I_{cb}|}{g^2}\right)}(\tau), \quad J \equiv (|I_{ca}|k-|I_{cb}|j+|I_{ca}I_{cb}|\ell_0)/g \in \mathbb{Z}_{\left(\frac{|I_{ab}I_{ca}I_{cb}|}{g^2}\right)},
\end{align}
where $c={\cal A}^{-1/2} \left|\frac{I_{ab}I_{ca}}{I_{cb}}\right|^{1/4}$ denotes the modular invariant overall factor which does not contribute to the flavor structure. 
Then, they transform as the modular forms of weight $1/2$~\footnote{Since we can find it from Eq.~(\ref{Normalization}) that each of the wavefunctions of matter fields as well as Higgs fields on the 4D space-time has modular weight $-k=-1/2$, (which is discussed in Ref.~\cite{Kikuchi:2020frp}), the superpotential in the 4D supergravity theory, $W$ has modular weight $-1$, which is consistent within the framework of the supergravity theory. See for studies on these aspects  in supergravity theory, e.g. Ref.~\cite{Kobayashi:2016ovu}.} for $\widetilde{\Gamma}(2|I_{ab}I_{ca}I_{cb}|/g^2)$~\footnote{It corresponds to $\mathbb{Z}_g$ shift invariant modes (at $z=0$) on the $T^2/\mathbb{Z}_g$ full shifted orbifold with the magnetic flux $|I_{ab}I_{ca}I_{cb}|$. See for such symmetry, e.g. \cite{Abe:2009vi}.}:
\begin{align}
  Y^{ijk}(\widetilde{\gamma}\tau) &= c\sum_{n=1}^g \eta_{g\left(J+(n-1)\frac{|I_{ab}I_{ca}I_{cb}|}{g^2}\right)}(\widetilde{\gamma}\tau) \notag \\
   &= \widetilde{J}_{1/2}(\widetilde{\gamma},\tau) \sum_{K=0}^{|I_{ab}I_{ca}I_{cb}|/g^2-1} \widetilde{\rho}^{(Y)}(\widetilde{\gamma})_{JK} c\sum_{n=1}^g \eta_{g\left(K+(n-1)\frac{|I_{ab}I_{ca}I_{cb}|}{g^2}\right)}(\tau),
\end{align}
where
\begin{align}
  \widetilde{\rho}^{(Y)}(\widetilde{S})_{JK} = e^{i\pi/4} \frac{g}{\sqrt{|I_{ab}I_{ca}I_{cb}|}} e^{2\pi i\left(\frac{g^2}{|I_{ab}I_{ca}I_{cb}|}\right)JK},\quad
  \widetilde{\rho}^{(Y)}(\widetilde{T})_{JK} = e^{i\pi \left(\frac{g^2}{|I_{ab}I_{ca}I_{cb}|}\right)J^2} \delta_{J,K}.
\end{align}
These unitary representations for the modular transformation are the same as  ones for
the zero-modes on $T^2$, Eq~(\ref{rhoSandT}), replacing $|I_{ab}I_{ca}I_{cb}|/g^2$ with $|I_{\alpha\beta}|$.
Thus, $\widetilde{\rho}^{(Y)}$ satisfies the algebraic relations for $\widetilde{\Gamma}_{2|I_{ab}I_{ca}I_{cb}|/g^2}$:
\begin{align}
  \begin{array}{c}
    \widetilde{\rho}(\widetilde{S})^2 = \widetilde{\rho}(\widetilde{Z}),\ \widetilde{\rho}(\widetilde{S})^4 = [\widetilde{\rho}(\widetilde{S})\widetilde{\rho}(\widetilde{T})]^3 = -\widetilde{\mathbb{I}}, \ 
    \widetilde{\rho}(\widetilde{Z})\widetilde{\rho}(\widetilde{T}) = \widetilde{\rho}(\widetilde{T})\widetilde{\rho}(\widetilde{Z}),\ \widetilde{\rho}(\widetilde{T})^{2|I_{ab}I_{ca}I_{cb}|/g^2} = \widetilde{\mathbb{I}}, \label{YukawaAlgebra}
  \end{array}
\end{align}
where we omit the symbol $(Y)$.
These algebraic relations can be also derived from the tensor products of the representations.
Under modular transformation, the Yukawa coupling $\widetilde{Y}^{ijk}$ in Eq.~(\ref{YukawaonT2}) as well as the holomorphic Yukawa coupling $Y^{ijk}$ in Eq.~(\ref{holomY}) is transformed by the following unitary representation: 
\begin{align}
  \widetilde{\rho}^{(L\otimes R \otimes H^*)}(\widetilde{\gamma})_{(ijk)(i'j'k')} = \widetilde{\rho}_{T^2}^{(ab)}(\widetilde{\gamma})_{ii'} \otimes \widetilde{\rho}_{T^2}^{(ca)}(\widetilde{\gamma})_{jj'} \otimes \left(\widetilde{\rho}^{(cb)}_{T^2}(\widetilde{\gamma})_{kk'} \right)^*,
\end{align}
where $\widetilde{\rho}_{T^2}^{(\alpha\beta)}$ is the unitary representation for the modular transformation acting on the $\alpha\beta$-sector
zero-modes satisfying the algebraic relarions for $\widetilde{\Gamma}_{2|I_{\alpha\beta}|}$ in Eq.~(\ref{AlgebraonT2}) with $\alpha\beta \in \{ab,ca,cb\}$.
Thus, $\widetilde{\rho}^{(L\otimes R \otimes H^*)}$ should satisfy the following algebraic relations:
\begin{align}
  \begin{array}{c}
    \widetilde{\rho}(\widetilde{S})^2 = \widetilde{\rho}(\widetilde{Z}),\ \widetilde{\rho}(\widetilde{S})^4 = [\widetilde{\rho}(\widetilde{S})\widetilde{\rho}(\widetilde{T})]^3 = -\widetilde{\mathbb{I}}, \ 
    \widetilde{\rho}(\widetilde{Z})\widetilde{\rho}(\widetilde{T}) = \widetilde{\rho}(\widetilde{T})\widetilde{\rho}(\widetilde{Z}),\ \widetilde{\rho}(\widetilde{T})^{2|{\rm lcm}(I_{ab},I_{ca},I_{cb})|} = \widetilde{\mathbb{I}}, \label{LRHAlgebra}
  \end{array}
\end{align}
where we omit the symbol $(L\otimes R\otimes H^*)$.
We can check that the order of $\widetilde{T}$ in Eq.~(\ref{LRHAlgebra}), $2|{\rm lcm}(I_{ab},I_{ca},I_{cb})|$, is equivalent to one in Eq.~(\ref{YukawaAlgebra}), $2|I_{ab}I_{ca}I_{cb}|/g^2$, because of the fluxes relation in Eq.~(\ref{FluxConditionThree}).

It is straightforward to extend this result to the Yukawa couplings on $T^2_1 \times T^2_2$; we can show the holomorphic Yukawa couplings transform as the modular forms of weight 1 for $\Gamma(2{\rm lcm}(|I_{ab}^{(1)}I_{ca}^{(1)}I_{cb}^{(1)}|/g_1^2,|I_{ab}^{(2)}I_{ca}^{(2)}I_{cb}^{(2)}|/g^2_2))$ merely considering the tensor product representations of $\widetilde{\rho}(\widetilde{\gamma})$.
In Tables \ref{ThreeGenT}, \ref{ThreeGenP}, and \ref{ThreeGenTP},
we show the algebraic relations for the (holomorphic) Yukawa couplings on the three generation models we have classified in section \ref{ThreeonT2xT2}.


\subsection{An illustrating example: \\ 2-2-4 model on the $(T^2 \times T^2)/(\mathbb{Z}_2^{\rm (t)} \times \mathbb{Z}_2^{\rm (per)})$  orbifold}
\label{Ill224}

\subsubsection{Yukawa matrices}

In this subsection, we study the $(|{I}_{ab}^{(1)}|,|{I}_{ca}^{(1)}|,|{I}_{cb}^{(1)}|) = (|{I}_{ab}^{(2)}|,|{I}_{ca}^{(2)}|,|{I}_{cb}^{(2)}|) = (2,2,4)$ model on the $(T^2_1 \times T^2_2)/(\mathbb{Z}_2^{\rm (t)} \times \mathbb{Z}_2^{\rm (per)})$  orbifold shown in Table \ref{ThreeGenTP} as the model type of (T,P) = (I, I).
Thus, all zero-modes of left-handed matter fields $L^I$, right-handed matter fields $R^J$ and Higgs fields $H^K$ are even wavefunctions as shown in Table \ref{ModePatternTP}. 
We show their explicit forms in Table \ref{Zeromodeon224}.
There are six zero-modes for Higgs fields.

\begin{table}[H]
  \begin{center}
	\begingroup
	\renewcommand{\arraystretch}{1.3}
	$\begin{array}{c|c|c|c}
		& L^I (\lambda^{ab})& R^J (\lambda^{ca})& H^K(\lambda^{cb}) \\ \hline
		(00) & \psi^0_{T^2_1}\psi^0_{T^2_2} & \psi^0_{T^2_1}\psi^0_{T^2_2} & \psi^0_{T^2_1}\psi^0_{T^2_2} \\
		(10) & \frac{1}{\sqrt{2}}(\psi^1_{T^2_1}\psi^0_{T^2_2}+\psi^0_{T^2_1}\psi^1_{T^2_2}) & \frac{1}{\sqrt{2}}(\psi^1_{T^2_1}\psi^0_{T^2_2}+\psi^0_{T^2_1}\psi^1_{T^2_2}) & \frac{1}{2}(\psi^1_{T^2_1}+\psi^3_{T^2_1})\psi^0_{T^2_2}+\frac{1}{2}\psi^0_{T^2_1}(\psi^1_{T^2_2}+\psi^3_{T^2_2}) \\
		(11) & \psi^1_{T^2_1}\psi^1_{T^2_2} & \psi^1_{T^2_1}\psi^1_{T^2_2} & \frac{1}{2}(\psi^1_{T^2_1}+\psi^3_{T^2_1})(\psi^1_{T^2_2}+\psi^3_{T^2_2}) \\
		(20) & $-$ & $-$ & \frac{1}{\sqrt{2}}(\psi^2_{T^2_1}\psi^0_{T^2_2}+\psi^0_{T^2_1}\psi^2_{T^2_2}) \\
		(21) & $-$ & $-$ & \frac{1}{2}\psi^2_{T^2_1}(\psi^1_{T^2_2}+\psi^3_{T^2_2})+\frac{1}{2}(\psi^1_{T^2_1}+\psi^3_{T^2_1})\psi^2_{T^2_2} \\
		(22) & $-$ & $-$ & \psi^2_{T^2_1}\psi^2_{T^2_2} \\
	\end{array}$
	\endgroup
    \caption{Zero-mode wavefunctions in the 2-2-4 model on $(T^2_1\times T^2_2)/(\mathbb{Z}_2^{\rm (t)} \times \mathbb{Z}_2^{\rm (per)})$ orbifold.}
    \label{Zeromodeon224}
  \end{center}
\end{table}

Then, Yukawa couplings $Y^{IJK}H^K$ are written by
\begin{align}
  Y^{IJK}H^K = y_{IJ}^{(00)}H^{(00)} + y_{IJ}^{(10)}H^{(10)} + y_{IJ}^{(11)}H^{(11)} + y_{IJ}^{(20)}H^{(20)} + y_{IJ}^{(21)}H^{(21)} + y_{IJ}^{(22)}H^{(22)}, \notag
\end{align}
where
\begin{align}
  \begin{array}{cc}
    y^{(00)}_{IJ}=
    \begin{pmatrix}
      y_{a}&0&0\\
      0&y_{b}&0\\
      0&0&y_{c}\\
    \end{pmatrix},
    &
    y^{(10)}_{IJ}=
    \begin{pmatrix}
      0&y_{d}&0\\
      y_{d}&0&y_{e}\\
      0&y_{e}&0\\
    \end{pmatrix},\\
    y^{(11)}_{IJ}=
    \begin{pmatrix}
      0&0&y_{f}\\
      0&y_{f}&0\\
      y_{f}&0&0\\
    \end{pmatrix},
    &
    y^{(20)}_{IJ}=
    \begin{pmatrix}
      \sqrt{2}y_{b}&0&0\\
      0&\frac{1}{\sqrt{2}}(y_{a}+y_{c})&0\\
      0&0&\sqrt{2}y_{b}\\
    \end{pmatrix},\\
    y^{(21)}_{IJ}=
    \begin{pmatrix}
      0&y_{e}&0\\
      y_{e}&0&y_{d}\\
      0&y_{d}&0\\
    \end{pmatrix},
    &
    y^{(22)}_{IJ}=
    \begin{pmatrix}
      y_{c}&0&0\\
      0&y_{b}&0\\
      0&0&y_{a}\\
    \end{pmatrix},
  \end{array}
\end{align}
and
\begin{align}
  \begin{array}{ll}
    y_{a}=(\eta_0+\eta_8)^2, & y_{b}=2(\eta_0+\eta_8)\eta_4,\\
    y_{c}=4\eta_4^2, & y_{d}=\sqrt{2}(\eta_0+\eta_8)(\eta_2+\eta_6),\\
    y_{e}=2\sqrt{2}\eta_4(\eta_2+\eta_6), & y_{f}=2(\eta_2+\eta_6)^2.
  \end{array} \notag
\end{align}
Note that we have used the following property of the $\vartheta$ function,
\begin{align}
  \vartheta
  \begin{bmatrix}
    \frac{j}{M} \\ 0
  \end{bmatrix}
  (0,M\tau)
  =
  \vartheta
  \begin{bmatrix}
    \frac{M-j}{M} \\ 0
  \end{bmatrix}
  (0,M\tau), \quad j \in \mathbb{Z}_M, M \in \mathbb{Z},
\end{align}
and the notation $\eta_N$ defined in Eq.~(\ref{etanotation}) with $M = 16$.
Furthermore, in terms of modular symmetry, the zero-modes of left-handed matter fields, right-handed matter fields, and Higgs fields transform under $\Gamma_4'$ with weight 1, $\Gamma_4'$ with weight 1, $\Gamma_8'$ with weight 1, respectively\footnote{See Ref.\cite{Kikuchi:2020} for the detail of the group structure.}.
Also, the Yukawa couplings are the modular forms of weight 1 for $\Gamma(8)$ and then transform under $\Gamma_8'$.


\subsubsection{
Numerical analysis}
\label{Num224}

Now, we show the numerical studies  on the 2-2-4 model.
We assume that three zero-modes with $|{I}_{ab}^{(1)}|=|{I}_{ab}^{(2)}|=2$, three zero-modes with $|{I}_{ca}^{(1)}|=|{I}_{ca}^{(2)}|=2$, and six zero-modes with $|{I}_{cb}^{(1)}|=|{I}_{cb}^{(2)}|=4$ are regarded as three generation left-handed quark doublets, three generation right-handed quark doublets, and six generation up and down type Higgs fields
under the gauge group broken from $U(8)$,
respectively. Then, both Yukawa matrices for up and down-sectors as ones in the 2-2-4 model on the $(T^2_1 \times T^2_2)/(\mathbb{Z}_2^{\rm (t)} \times \mathbb{Z}_2^{\rm (per)})$  orbifold.
We also assume the vacuum expectation values (VEVs) of the up and down-sectors of the Higgs fields are independent; we can realize the quark mixing 
taking the different vacuum alignments of the Higgs fields
for the up and down-sectors.

First, we show the simplest example.
When we choose the following vacuum alignments of Higgs fields,
\begin{align}
  \begin{array}{c}
    \langle H_u^{(22)} \rangle \neq 0, \ ({\rm other\ VEVs})=0, \\
    \langle H_d^{(21)} \rangle = 0.4 \langle H_d^{(22)} \rangle \neq 0, \ ({\rm other\ VEVs})=0,
  \end{array}
  \label{VEVs1}
\end{align}
the Yukawa matrices can be written as
\begin{align}
  Y^{IJK}_u H_K = 
  \begin{pmatrix}
    y_c & & \\
    & y_b & \\
    & & y_a 
  \end{pmatrix}
  H_u^{(22)},
\end{align}
\begin{align}
  Y^{IJK}_d H_K = 
  \begin{pmatrix}
    y_c & 0.4y_e & 0 \\
    0.4 y_e & y_b & 0.4y_d \\
    0 & 0.4y_d & y_a 
  \end{pmatrix}
  H_d^{(22)}.
\end{align}
For $\tau = 1.5i$, we can obtain the mass ratios of the quarks and the absolute values of the CKM matrix elements as shown in Table \ref{MassandCKM1}.

\begin{table}[H]
  \begin{center}
    \renewcommand{\arraystretch}{1.2}
    $\begin{array}{c||c|c} \hline
      & {\rm Obtained\ values} & {\rm Observed\ values} \\ \hline
      (m_u,m_c,m_t)/m_t & (3.22\times 10^{-4},1.80\times 10^{-2},1) & (1.26\times 10^{-5},7.38\times 10^{-3},1) \\ \hline
      (m_d,m_s,m_b)/m_b & (1.02\times 10^{-3},1.24\times 10^{-2},1) & (1.12\times 10^{-3},2.22\times 10^{-2},1) \\ \hline
      |V_{\rm CKM}| \equiv |{(U_L^u)}^{\dagger}U_L^d|
      &
      \begin{pmatrix}
        0.973 & 0.230 & 0.000515 \\
        0.226 & 0.959 & 0.170 \\
        0.0395 & 0.165 & 0.986 
      \end{pmatrix}
      & 
      \begin{pmatrix}
        0.974 & 0.227 & 0.00361 \\
        0.226 & 0.973 & 0.0405 \\
        0.00854 & 0.0398 & 0.999 
      \end{pmatrix}\\ \hline
    \end{array}$
    \caption{The mass ratios of the quarks and the absolute values of the CKM matrix elements at $\tau=1.5i$ under the vacuum alignments of Higgs fields in Eq.~(\ref{VEVs1}).
    Observed values are shown in Ref \cite{Zyla:2020zbs}.}
    \label{MassandCKM1}
  \end{center}
\end{table}

Assuming further non-vanishing VEVs of Higgs fields, we can realize more realistic results.
For example, when we choose the following vacuum alignments of Higgs fields,
\begin{align}
  \begin{array}{c}
    \langle H_u^{(21)} \rangle = 0.22 \langle H_u^{(22)} \rangle \neq 0, \ ({\rm other\ VEVs})=0, \\
    \langle H_d^{(20)} \rangle = -0.10 \langle H_d^{(22)} \rangle \neq 0,\ \langle H_d^{(21)} \rangle = 0.34 \langle H_d^{(22)} \rangle \neq 0, \ ({\rm other\ VEVs})=0,
  \end{array}
  \label{VEVs2}
\end{align}
the Yukawa matrices can be written as
\begin{align}
  Y^{IJK}_u H_K = 
  \begin{pmatrix}
    y_c & 0.22y_e & 0 \\
    0.22y_e & y_b & 0.22y_d \\
    0 & 0.22y_d & y_a 
  \end{pmatrix}
  H_u^{(22)},
\end{align}
\begin{align}
  Y^{IJK}_d H_K = 
  \begin{pmatrix}
    y_c-0.10\times \sqrt{2}y_b & 0.34y_e & 0 \\
    0.34 y_e & y_b-\frac{0.10}{\sqrt{2}}(y_a+y_c) & 0.34y_d \\
    0 & 0.34y_d & y_a-0.10 \times \sqrt{2}y_b 
  \end{pmatrix}
  H_d^{(22)}.
\end{align}
For $\tau = 1.5i$, we can obtain the mass ratios of the quarks and the absolute values of the CKM matrix elements as shown in Table \ref{MassandCKM2}.

\begin{table}[H]
  \begin{center}
  \renewcommand{\arraystretch}{1.2}
    $\begin{array}{c||c|c} \hline
      & {\rm Obtained\ values} & {\rm Observed\ values} \\ \hline
      (m_u,m_c,m_t)/m_t & (1.35\times 10^{-5},8.96\times 10^{-3},1) & (1.26\times 10^{-5},7.38\times 10^{-3},1) \\ \hline
      (m_d,m_s,m_b)/m_b & (2.08\times 10^{-3},7.20\times 10^{-2},1) & (1.12\times 10^{-3},2.22\times 10^{-2},1) \\ \hline
      |V_{\rm CKM}| \equiv |{(U_L^u)}^{\dagger}U_L^d|
      &
      \begin{pmatrix}
        0.973 & 0.229 & 0.00772 \\
        0.229 & 0.973 & 0.0403 \\
        0.00171 & 0.0410 & 0.999 
      \end{pmatrix}
      & 
      \begin{pmatrix}
        0.974 & 0.227 & 0.00361 \\
        0.226 & 0.973 & 0.0405 \\
        0.00854 & 0.0398 & 0.999 
      \end{pmatrix}\\ \hline
    \end{array}$
    \caption{The mass ratios of the quarks and the absolute values of the CKM matrix elements at $\tau=1.5i$ under the vacuum alignments of Higgs fields in Eq.~(\ref{VEVs2}).
Observed values are shown in Ref \cite{Zyla:2020zbs}.}
    \label{MassandCKM2}
  \end{center}
\end{table}

As a result of our model, we can realize realistic values of quark mass ratios and mixing angles 
by assuming proper Higgs VEV alignment and a proper value of $\tau$.
Similarly we can study other orbifold models.

\section{Conclusion}
\label{conclusion}

We have studied the $\mathbb{Z}_2^{\rm (per)}$ permutaion orbifold of  $T^2_1 \times T^2_2$ and its twisted orbifolds.
We have classified the possible three generation models with non-vanishing Yukawa couplings based on the magnetized 
$(T^2_1 \times T^2_2)/\mathbb{Z}_2^{\rm (per)}$ orbifold and $(T^2_1\times T^2_2)/(\mathbb{Z}_2^{\rm (t)} \times \mathbb{Z}_2^{\rm (per)})$  orbifold as well as $(T^2_1 \times T^2_2)/\mathbb{Z}_2^{\rm (t)}$. 
Consequently, choosing the suitable fluxes, there are many possibilities as shown in section \ref{ThreeonT2xT2}.
We have a rich structure in model building and various flavor structures.
Also, we have studied the modular symmetry of the Yukawa couplings on $T^2$ which characterizes the values of Yukawa matrix elements.
Under the modular transformation, the holomorphic part of Yukawa couplings, $Y^{ijk}(\tau)$, transforms as the modular form of weight 1/2 for $\widetilde{\Gamma}(2N)$, $N\equiv |{\rm lcm}(I_{ab},I_{ca},I_{cb})|=|I_{ab}I_{ca}I_{cb}|/g^2$.
Simiraly, one on $T^2_1 \times T^2_2$, $Y^{IJK}(\tau)$, transforms as the modular form of weight 1 for $\Gamma(2{\rm lcm}(N_1,N_2))$.

As an illustrating example, we have investigated the 2-2-4 model on $(T^2_1 \times T^2_2)/(\mathbb{Z}_2^{\rm (t)} \times \mathbb{Z}_2^{\rm (per)})$ orbifold whose Yukawa coupling $Y^{IJK}(\tau)$ is the modular form of weight 1 for $\Gamma(8)$.
Then, the realistic mass ratios of quarks and values of the CKM matrix elements can be realized well by adjusting the vaccum alignments of Higgs fields and the complex structure modulus as free parameters.

We can extend our analysis to other three generation models including the existences of leptons.
We need the moduli stabilization to fix $\tau$.
For example, in Ref.~\cite{Ishiguro:2020tmo}, it was found that certain values of $\tau$ are favorable from the 
viewpoint of moduli stabilization.
It is interesting to apply those results in numerical analyses of our orbifold models.

We need to study mass matrices of Higgs fields to fix the Higgs VEV direction.
There is no Higgs mass term, the so-called $\mu$-term at the tree level.
The $\mu$-term matrices can be induced by non-perturbative effects such as 
D-brane instant effects \cite{Blumenhagen:2006xt,Ibanez:2006da,Ibanez:2007rs,Antusch:2007jd,Kobayashi:2015siy}.
This issue is important, but beyond our scope.
We will study it elsewhere.

Similarly, we can study charged lepton masses as well as Dirac neutrino masses.
However, we also need to study Majorana masses in order to understand the neutrino masses and the lepton mixing anlges.
Majorana masses are also vanishing at the tree level, but induced by non-perturbative effects such as 
D-brane instant effects.
We will also study it elsewhere.

\vspace{1.5 cm}
\noindent
{\large\bf Acknowledgement}\\

T. K. was supported in part by MEXT KAKENHI Grant Number JP19H04605. 
H. U. was supported by Grant-in-Aid for JSPS Research Fellows No. 20J20388.


\appendix

\section{
Magnetized $T^2/\mathbb{Z}_N$ shifted orbifold models}
\label{T2/ZN shifted orbifolds}
Here, we study the $T^2/\mathbb{Z}_N$ shifted orbifolds.
The $\mathbb{Z}_N$ shift is defined by the following transformation of the complex coordinate on $T^2$ \cite{Fujimoto:2013xha}:
\begin{align}
  z \rightarrow z + ke_N^{(m,n)}, \label{shift}
\end{align}
where
\begin{align}
  e_N^{(m,n)} \equiv (m+n\tau)/N; ^\forall k, ^\exists m, ^\exists n \in \mathbb{Z}_N.
\end{align}
To preserve the full modular symmetry, it is required to identify $z$ with $z + ke_N^{(m,n)}$ for all $m,n\in\mathbb{Z}_N$.
Then, we consider the eigenmode
under full $\mathbb{Z}_N$ shifts, that is, $T^2/\mathbb{Z}_N$ full shifted orbifolds, for zero-modes.
Any $\mathbb{Z}_N$ shift is generated by the following two shifts:
\begin{align}
  e_N^{(1,0)} = 1/N,\quad e_N^{(0,1)} = \tau/N.
\end{align}
The zero-modes
on $T^2$ are transformed as
\begin{align}
  &\psi^{j,|I_{\alpha\beta}|}(z) \rightarrow \psi^{j,|I_{\alpha\beta}|}(z+ke_N^{(m,n)})
  = e^{ik\chi_N^{(m,n)}(z)} e^{i\pi km(2j-(N-k)n|I_{\alpha\beta}|/N)/N} \psi^{j+kn|I_{\alpha\beta}|/N,|I_{\alpha\beta}|}(z),
\end{align}
where
\begin{align}
  \chi_N^{(m,n)}(z) \equiv \pi |I_{\alpha\beta}|\left( \frac{{\rm Im}(\bar{e}_N^{(m,n)}(z+\tau))}{{\rm Im}\tau}+\frac{mn}{N} \right).
\end{align}
The zero-modes
on the $T^2/\mathbb{Z}_N$ full shifted orbifolds, $\Psi^{r,|s_{\alpha\beta}|}_{T^2/\mathbb{Z}_N^{(\ell_1,\ell_2)}}(z)$, are required to obey the following boundary condition,
\begin{align}
  \Psi^{r,|s_{\alpha\beta}|}_{T^2/\mathbb{Z}_N^{(\ell_1,\ell_2)}}(z+ke_N^{(m,n)}) = e^{2\pi ik\ell^{(m,n)}/N} e^{i\chi_N^{(m,n)}(z)} \Psi^{r,|s_{\alpha\beta}|}_{T^2/\mathbb{Z}_N^{(\ell_1,\ell_2)}}(z),
  \quad \ell_1, \ell_2 \in \mathbb{Z}_N,
\end{align}
with the magnetic flux condition $|I_{\alpha\beta}|/N^2 \equiv |s_{\alpha\beta}| \in \mathbb{Z}$ and the phase conditions:
\begin{align}
  \left\{
  \begin{array}{l}
    \ell^{(m,n)} = m\ell_1 + n\ell_2\ ({\rm mod}\ N)\quad {\rm for}\ N\in\mathbb{Z},|s_{\alpha\beta}| \in 2\mathbb{Z} \\
    \ell^{(m,n)} = m\ell_1 + n\ell_2\ ({\rm mod}\ N)\quad {\rm for}\ N\in2\mathbb{Z}+1,|s_{\alpha\beta}| \in 2\mathbb{Z}+1 \\
    \ell^{(m,n)} = m\ell_1 + n\ell_2+mnN/2\ ({\rm mod}\ N)\quad {\rm for}\ N\in2\mathbb{Z},|s_{\alpha\beta}| \in 2\mathbb{Z}+1
  \end{array}
  \right..
\end{align}
Thus, they can be expanded by the
zero-modes on $T^2$ as
\begin{align}
  \Psi^{r,|s_{\alpha\beta}|}_{T^2/\mathbb{Z}_N^{(\ell_1,\ell_2)}}(z) &= \frac{1}{\sqrt{N}} \sum_{k=0}^{N-1} e^{-2\pi ik\ell_2/N} \psi^{j+kN|s_{\alpha\beta}|,|I_{\alpha\beta}|}(z), \label{psishift} \\
  j=Nr+\ell_1 &\in \mathbb{Z}_{N|s_{\alpha\beta}|},\ r \in \mathbb{Z}_{|s_{\alpha\beta}|},\ \ell_1,\ell_2 \in \mathbb{Z}_N. \notag
\end{align}
The number of zero-modes for any $(\ell_1,\ell_2)$, $\ell_1, \ell_2 \in \mathbb{Z}_N$ on the $T^2/\mathbb{Z}_N$ full shifted orbifolds equals to $|I_{\alpha\beta}|/N^2 = |s_{\alpha\beta}|$.
This is related to the case on $T^2$ with the magnetic flux $|s_{\alpha\beta}|$ considering $\mathbb{Z}_N$ Scherk-Schwarz phases\footnote{See in detail Ref.\cite{Kikuchi:2020}.}.
In particular, the $\mathbb{Z}_N$ shift invariant modes, $(\ell_1,\ell_2)=(0,0)$, for $|s_{\alpha\beta}| \in 2\mathbb{Z}$ transform as the modular forms of weight $1/2$ for $\widetilde{\Gamma}(2|s_{\alpha\beta}|)$ under the modular transformation which corresponds to Eqs.~(\ref{wavemodularform}) and (\ref{rhoSandT}).



\begin{thebibliography}{99}



\bibitem{Bachas:1995ik}
  C.~Bachas,
  arXiv:hep-th/9503030.

\bibitem{Blumenhagen:2000wh}
  R.~Blumenhagen, L.~Goerlich, B.~Kors and D.~Lust,
  JHEP {\bf 0010}, 006 (2000)
  [arXiv:hep-th/0007024].

\bibitem{Angelantonj:2000hi}
  C.~Angelantonj, I.~Antoniadis, E.~Dudas and A.~Sagnotti,
  Phys. Lett. {\bf B489}, 223 (2000)
  [arXiv:hep-th/0007090].

\bibitem{Blumenhagen:2000ea}
  R.~Blumenhagen, B.~Kors and D.~Lust,
  JHEP {\bf 0102}, 030 (2001)
  [arXiv:hep-th/0012156].

\bibitem{Cremades:2004wa}
D.~Cremades, L.~E.~Ibanez and F.~Marchesano,
JHEP \textbf{05} (2004), 079
[arXiv:hep-th/0404229 [hep-th]].


\bibitem{Kobayashi:2018rad} 
 T.~Kobayashi, S.~Nagamoto, S.~Takada, S.~Tamba and T.~H.~Tatsuishi,
 Phys.\ Rev.\ D {\bf 97}, no. 11, 116002 (2018)
 [arXiv:1804.06644 [hep-th]].



\bibitem{Ohki:2020bpo}
H.~Ohki, S.~Uemura and R.~Watanabe,
Phys. Rev. D \textbf{102}, no.8, 085008 (2020)
[arXiv:2003.04174 [hep-th]].



\bibitem{Kikuchi:2020frp}
S.~Kikuchi, T.~Kobayashi, S.~Takada, T.~H.~Tatsuishi and H.~Uchida,
Phys. Rev. D \textbf{102}, no.10, 105010 (2020)
[arXiv:2005.12642 [hep-th]].




\bibitem{Kikuchi:2020nxn}
S.~Kikuchi, T.~Kobayashi, H.~Otsuka, S.~Takada and H.~Uchida,
[arXiv:2007.06188 [hep-th]].

\bibitem{Kobayashi:2018bff}
T.~Kobayashi and S.~Tamba,
Phys.\ Rev.\ D {\bf 99} (2019) no.4, 046001
[arXiv:1811.11384 [hep-th]].




\bibitem{Feruglio:2017spp} 
  F.~Feruglio,
  arXiv:1706.08749 [hep-ph];
%
  T.~Kobayashi, K.~Tanaka and T.~H.~Tatsuishi,
  Phys.\ Rev.\ D {\bf 98}, no. 1, 016004 (2018)
  [arXiv:1803.10391 [hep-ph]];
%
  J.~T.~Penedo and S.~T.~Petcov,
  Nucl.\ Phys.\ B {\bf 939}, 292 (2019)
  [arXiv:1806.11040 [hep-ph]];
%
  J.~C.~Criado and F.~Feruglio,
  SciPost Phys.\  {\bf 5}, no. 5, 042 (2018)
  [arXiv:1807.01125 [hep-ph]];
%
  T.~Kobayashi, N.~Omoto, Y.~Shimizu, K.~Takagi, M.~Tanimoto and T.~H.~Tatsuishi,
  JHEP {\bf 1811}, 196 (2018)
  [arXiv:1808.03012 [hep-ph]];
%
  P.~P.~Novichkov, J.~T.~Penedo, S.~T.~Petcov and A.~V.~Titov,
  JHEP {\bf 1904}, 005 (2019)
  [arXiv:1811.04933 [hep-ph]];
%
  JHEP {\bf 1904}, 174 (2019)
  [arXiv:1812.02158 [hep-ph]];
%
  F.~J.~de Anda, S.~F.~King and E.~Perdomo,
  arXiv:1812.05620 [hep-ph];
%
  H.~Okada and M.~Tanimoto,
  Phys.\ Lett.\ B {\bf 791}, 54 (2019)
  [arXiv:1812.09677 [hep-ph]];
%
  T.~Kobayashi, Y.~Shimizu, K.~Takagi, M.~Tanimoto, T.~H.~Tatsuishi and H.~Uchida,
  Phys.\ Lett.\ B {\bf 794}, 114 (2019)
  [arXiv:1812.11072 [hep-ph]];
%
  P.~P.~Novichkov, S.~T.~Petcov and M.~Tanimoto,
  Phys.\ Lett.\ B {\bf 793}, 247 (2019)
  [arXiv:1812.11289 [hep-ph]].



\bibitem{Baur:2019kwi}
  A.~Baur, H.~P.~Nilles, A.~Trautner and P.~K.~S.~Vaudrevange,
  Phys.\ Lett.\ B {\bf 795} (2019) 7
  [arXiv:1901.03251 [hep-th]];
%
%
Nucl. Phys. B \textbf{947} (2019), 114737
[arXiv:1908.00805 [hep-th]].  


\bibitem{Nilles:2020tdp}
H.~P.~Nilles, S.~Ramos\textendash{}S\'anchez and P.~K.~S.~Vaudrevange,
Phys. Lett. B \textbf{808} (2020), 135615
[arXiv:2006.03059 [hep-th]];
%
%
[arXiv:2010.13798 [hep-th]].


\bibitem{Dixon:1985jw}
L.~J.~Dixon, J.~A.~Harvey, C.~Vafa and E.~Witten,
Nucl. Phys. B \textbf{261}, 678-686 (1985);
Nucl. Phys. B \textbf{274}, 285-314 (1986).



\bibitem{Abe:2008fi}
H.~Abe, T.~Kobayashi and H.~Ohki,
JHEP \textbf{09} (2008), 043
[arXiv:0806.4748 [hep-th]].


\bibitem{Abe:2013bca} 
  T.~H.~Abe, Y.~Fujimoto, T.~Kobayashi, T.~Miura, K.~Nishiwaki and M.~Sakamoto,
  JHEP {\bf 1401}, 065 (2014)
  [arXiv:1309.4925 [hep-th]].
 
  
\bibitem{Abe:2014noa} 
  T.~h.~Abe, Y.~Fujimoto, T.~Kobayashi, T.~Miura, K.~Nishiwaki and M.~Sakamoto,
  Nucl.\ Phys.\ B {\bf 890}, 442 (2014)
  [arXiv:1409.5421 [hep-th]].
 




\bibitem{Kobayashi:2017dyu} 
 T.~Kobayashi and S.~Nagamoto,
 Phys.\ Rev.\ D {\bf 96}, no. 9, 096011 (2017)
 [arXiv:1709.09784 [hep-th]].


\bibitem{Sakamoto:2020pev}
M.~Sakamoto, M.~Takeuchi and Y.~Tatsuta,
Phys. Rev. D \textbf{102}, no.2, 025008 (2020)
[arXiv:2004.05570 [hep-th]]; 
[arXiv:2010.14214 [hep-th]].



\bibitem{Abe:2008sx}
H.~Abe, K.~S.~Choi, T.~Kobayashi and H.~Ohki,
Nucl. Phys. B \textbf{814}, 265-292 (2009)
[arXiv:0812.3534 [hep-th]].


\bibitem{Abe:2015yva}
T.~h.~Abe, Y.~Fujimoto, T.~Kobayashi, T.~Miura, K.~Nishiwaki, M.~Sakamoto and Y.~Tatsuta,
Nucl. Phys. B \textbf{894}, 374-406 (2015)
[arXiv:1501.02787 [hep-ph]].



\bibitem{Abe:2012fj}
H.~Abe, T.~Kobayashi, H.~Ohki, A.~Oikawa and K.~Sumita,
Nucl. Phys. B \textbf{870}, 30-54 (2013)
[arXiv:1211.4317 [hep-ph]].  

\bibitem{Abe:2014vza}
H.~Abe, T.~Kobayashi, K.~Sumita and Y.~Tatsuta,
Phys. Rev. D \textbf{90}, no.10, 105006 (2014)
[arXiv:1405.5012 [hep-ph]].





\bibitem{Fujimoto:2016zjs}
Y.~Fujimoto, T.~Kobayashi, K.~Nishiwaki, M.~Sakamoto and Y.~Tatsuta,
Phys. Rev. D \textbf{94}, no.3, 035031 (2016)
[arXiv:1605.00140 [hep-ph]].


\bibitem{Abe:2016eyh}
H.~Abe, T.~Kobayashi, H.~Otsuka, Y.~Takano and T.~H.~Tatsuishi,
Phys. Rev. D \textbf{94}, no.12, 126020 (2016)
[arXiv:1605.00898 [hep-ph]].


\bibitem{Kobayashi:2016qag}
T.~Kobayashi, K.~Nishiwaki and Y.~Tatsuta,
JHEP \textbf{04}, 080 (2017)
[arXiv:1609.08608 [hep-th]].





\bibitem{Fujimoto:2013xha}
Y.~Fujimoto, T.~Kobayashi, T.~Miura, K.~Nishiwaki and M.~Sakamoto,
Phys.\ Rev.\ D \textbf{87} (2013) no.8, 086001
[arXiv:1302.5768 [hep-th]].



\bibitem{Antoniadis:2009bg}
I.~Antoniadis, A.~Kumar and B.~Panda,
Nucl. Phys. B \textbf{823} (2009), 116-173
[arXiv:0904.0910 [hep-th]].


\bibitem{Abe:2013bba}
H.~Abe, T.~Kobayashi, H.~Ohki, K.~Sumita and Y.~Tatsuta,
JHEP \textbf{04} (2014), 007
[arXiv:1307.1831 [hep-th]];
%
JHEP \textbf{06} (2014), 017
[arXiv:1404.0137 [hep-th]].

\bibitem{Abe:2009dr}
H.~Abe, K.~S.~Choi, T.~Kobayashi and H.~Ohki,
JHEP \textbf{06} (2009), 080
[arXiv:0903.3800 [hep-th]].


	\bibitem{Gunning:1962}
 	R. C. Gunning,
 	\textit{Lectures on Modular Forms}
 	(Princeton University Press, Princeton, NJ, 1962).

\bibitem{Schoeneberg:1974}
 B.~Schoeneberg,
  \textit{Elliptic Modular Functions}
  (Springer-Verlag, 1974)

\bibitem{Koblitz:1984}
 N.~Koblitz,
  \textit{Introduction to Elliptic Curves and Modular Forms}
  (Springer-Verlag, 1984)

\bibitem{Bruinier:2008}
J.H.~Bruinier, G.V.D.~Geer, G.~Harder, and D.~Zagier,
\textit{The 1-2-3 of Modular Forms}
(Springer, 2008)

\bibitem{deAdelhartToorop:2011re} 
R.~de Adelhart Toorop, F.~Feruglio and C.~Hagedorn,
Nucl.\ Phys.\ B {\bf 858}, 437 (2012)
[arXiv:1112.1340 [hep-ph]].


\bibitem{Liu:2019khw}
X.~G.~Liu and G.~J.~Ding,
JHEP {\bf 1908}, 134 (2019)
[arXiv:1907.01488 [hep-ph]];
%
P.~P.~Novichkov, J.~T.~Penedo and S.~T.~Petcov,
[arXiv:2006.03058 [hep-ph]];
%
X.~G.~Liu, C.~Y.~Yao and G.~J.~Ding,
[arXiv:2006.10722 [hep-ph]];
%
X.~G.~Liu, C.~Y.~Yao, B.~Y.~Qu and G.~J.~Ding,
[arXiv:2007.13706 [hep-ph]];
%
X.~Wang, B.~Yu and S.~Zhou,
[arXiv:2010.10159 [hep-ph]];
%
C.~Y.~Yao, X.~G.~Liu and G.~J.~Ding,
[arXiv:2011.03501 [hep-ph]].




\bibitem{shimura}
G.~Shimura,
Annals of Mathematics, 97(3), second series, 440 (1973)

\bibitem{Duncan:2018wbw}
J.~F.~Duncan and D.~A.~Mcgady,
[arXiv:1806.09875 [math.NT]].

\bibitem{Abe:2017gye}
H.~Abe, T.~Kobayashi, K.~Sumita and S.~Uemura,
Phys. Rev. D \textbf{96}, no.2, 026019 (2017)
[arXiv:1703.03402 [hep-th]].

\bibitem{Kobayashi:2016ovu} 
  T.~Kobayashi, S.~Nagamoto and S.~Uemura,
  PTEP {\bf 2017}, no. 2, 023B02 (2017)
  [arXiv:1608.06129 [hep-th]].

\bibitem{Abe:2009vi}
H.~Abe, K.~S.~Choi, T.~Kobayashi and H.~Ohki,
Nucl. Phys. B \textbf{820} (2009), 317-333
[arXiv:0904.2631 [hep-ph]].



\bibitem{Kikuchi:2020}
 S.~Kikuchi, T.~Kobayashi and H.~Uchida,
in preparation




\bibitem{Zyla:2020zbs}
P.~A.~Zyla \textit{et al.} [Particle Data Group],
PTEP \textbf{2020} (2020) no.8, 083C01


\bibitem{Ishiguro:2020tmo}
K.~Ishiguro, T.~Kobayashi and H.~Otsuka,
[arXiv:2011.09154 [hep-ph]].



\bibitem{Blumenhagen:2006xt}
R.~Blumenhagen, M.~Cvetic and T.~Weigand,
Nucl. Phys. B \textbf{771}, 113-142 (2007)
[arXiv:hep-th/0609191 [hep-th]].


\bibitem{Ibanez:2006da}
L.~E.~Ibanez and A.~M.~Uranga,
JHEP \textbf{03}, 052 (2007)
[arXiv:hep-th/0609213 [hep-th]].

\bibitem{Ibanez:2007rs}
L.~E.~Ibanez, A.~N.~Schellekens and A.~M.~Uranga,
JHEP \textbf{06}, 011 (2007)
[arXiv:0704.1079 [hep-th]].

\bibitem{Antusch:2007jd}
S.~Antusch, L.~E.~Ibanez and T.~Macri,
JHEP \textbf{09}, 087 (2007)
[arXiv:0706.2132 [hep-ph]].

\bibitem{Kobayashi:2015siy}
T.~Kobayashi, Y.~Tatsuta and S.~Uemura,
Phys. Rev. D \textbf{93}, no.6, 065029 (2016)
[arXiv:1511.09256 [hep-ph]].


\end{thebibliography}
\end{document}